\documentclass{aa}  

\usepackage{amsmath,amsfonts, amssymb}
\usepackage{graphicx,color}
\usepackage[dvipsnames,table]{xcolor}
\usepackage{txfonts}
\usepackage{longtable,booktabs}
\usepackage{xspace}
\usepackage{lscape}

\newcommand{\hi}{\textsc{H$\,$i}\xspace}
\newcommand{\numunit}[2]{\mbox{\ensuremath{#1\,#2}\xspace}}

\newcommand{\mhi}{\ensuremath{\text{M}_\hi}\xspace}
\newcommand{\kms}{\ensuremath{\text{km}\,\text{s}^{-1}}\xspace}
\newcommand{\msun}{\ensuremath{\text{M}_\odot}\xspace}
\newcommand{\mk}{MeerKAT\xspace}

\newcommand{\rvir}{\ensuremath{R_{200}}\xspace}
\newcommand{\caracal}{\texttt{CARAcal}\xspace}
\newcommand{\gipsy}{\texttt{GIPSY}\xspace}
\newcommand{\mhz}{\text{MHz}\xspace}
\newcommand{\mpc}{\text{Mpc}\xspace}
\newcommand{\pybdsf}{\texttt{PyBDSF}\xspace}
\newcommand{\sofia}{\texttt{SoFiA}\xspace}
\newcommand{\wfifty}{\ensuremath{\text{W}_{50}}\xspace}
\newcommand{\wtwenty}{\ensuremath{\text{W}_{20}}\xspace}

\newcommand{\hwone}{The Swarm\xspace}

\newcommand{\secref}[1]{Sec.~\ref{#1}}
\newcommand{\figref}[1]{Fig.~\ref{#1}}
\newcommand{\tabref}[1]{Table~\ref{#1}}

\begin{document}

   \title{MeerKAT 21-cm H$\,${\sc{i}} imaging of Abell 2626 and beyond}%
   \titlerunning{H$\,${\sc{i}} in Abell 2626 and beyond}
   \authorrunning{Healy-Deb et al.}

   \subtitle{}

   \author{J. Healy,\inst{1,2} \thanks{These authors contributed equally to this work}  \fnmsep  \thanks{healy@astro.rug.nl} % 0000-0003-1020-8684
          T. Deb,\inst{1}$^{\star}$  \fnmsep \thanks{deb@astro.rug.nl}
          M.A.W. Verheijen,\inst{1} % 0000-0001-9022-8081
          S-L. Blyth,\inst{2} % 0000-0002-5777-0036
          P. Serra,\inst{3} %orcid: 0000-0001-5965-252X
          M. Ramatsoku\inst{4,3} %orcid: 0000-0003-0231-3249
          \and
          B. Vulcani\inst{5} % orcid:  0000-0003-0980-1499
          }

  \institute{Kapteyn Astronomical Institute, University of Groningen, Landleven 12, 9747 AV Groningen, The Netherlands
         \and
         Department of Astronomy, University of Cape Town, Private Bag X3, Rondebosch 7701, South Africa
         \and
         INAF - Osservatorio Astronomico di Cagliari, Via della Scienza 5, I-09047 Selargius (CA), Italy
         \and
         Department of Physics and Electronics, Rhodes University, PO Box 94, Makhanda, 6140, South Africa
         \and
         INAF - Osservatorio astronomico di Padova, Vicolo Osservatorio 5, 35122 Padova, Italy
             }

   \date{Received 25 May, 2021; accepted June 22, 2021}

% \abstract{}{}{}{}{} 
% 5 {} token are mandatory
 
  \abstract
  {The morphology-density relation manifests the environmental dependence of the formation and evolution of galaxies as they continuously migrate through the cosmic web to ever denser environments. As gas-rich galaxies traverse the outskirts and inner regions of galaxy clusters they experience sudden and radical changes in their gas content and star formation activity.}
  {The goal of this work is to gain an \hi perspective on gas depletion mechanisms acting on galaxies and galaxy groups that are being accreted by a moderately massive galaxy cluster. We aim to study the relative importance and efficiency of processes such as ram-pressure stripping and tidal interactions as well as their dependency on the local and global environment of galaxies in the cluster core and in its surroundings.}
  {We have conducted a blind radio continuum and \hi spectral line imaging survey with the MeerKAT radio telescope of a 2$^{\circ}$ $\times$ 2$^{\circ}$ area centred on the galaxy cluster Abell 2626. We have used the \caracal pipeline to reduce the data, \sofia to detect sources within the \hi data cube, and \gipsy to construct spatially resolved information on the \hi morphologies and kinematics of the \hi detected galaxies.}
  {We have detected \hi in 219 galaxies with optical counterparts within the entire surveyed volume. We present the \hi properties of each of the detected galaxies as a data catalogue and as an atlas page for each galaxy, including \hi column-density maps, velocity fields, position-velocity diagrams and global \hi profiles. These data will also be used for case studies of identified ``jellyfish'' galaxies and galaxy population studies by means of morphological classification of the direct \hi detections as well as using the \hi stacking technique.}
   {}

   \keywords{galaxy clusters --
                galaxy evolution 
               }

   \maketitle

%-------------------------------------------------------------------
    
\section{Introduction}
    \label{sec:intro}

It is well known that galaxies inhabiting clusters differ from field galaxies, with ellipticals and lenticulars dominating in dense cluster environments and spirals prevailing in the field populations \citep{Hubble1931}. This observation is the so-called morphology-density relation \citep{Oemler1974, Dressler1980}. For galaxies of a fixed morphological type, cluster galaxies seem to have older stellar populations \citep{Bower1990, Rose1994, Cooper2010} and suppressed star formation (SF) \citep[e.g.][]{Balogh1997, Gomez2003, Boselli2020} relative to field galaxies. Furthermore, galaxies in clusters have a relatively smaller atomic gas reservoir compared to their field counterparts \citep{Haynes1984, Solanes2001, Gavazzi2005}. Since almost 40\% of galaxies in the universe reside in groups or clusters \citep[e.g.][]{Robotham2011}, it is clear that the cluster environment plays a crucial role in galaxy evolution in the universe.\\

Studies of clusters at intermediate redshifts ($z < 0.5$) show that both the morphologies and the SF rates in cluster galaxies significantly evolve over cosmic time \citep{Poggianti1999, Dressler1997, Fasano2000} with a decreasing fraction of lenticulars and an increasing fraction of spirals toward higher redshifts, while compact clusters at intermediate redshifts have a relatively low fraction of lenticulars with respect to ellipticals and more diffuse clusters have a relatively high fraction of lenticulars \citep{Fasano2000}. In addition, cluster galaxies at intermediate redshifts seem to display a higher star formation rate as evidenced by a larger fraction of blue galaxies \citep{Butcher1984}, known as the Butcher-Oemler effect, while \citet{Abraham1996} has shown that these blue galaxies are preferentially located in the infall region of clusters.\\

Thus, the dense environments of galaxy clusters somehow expedite the quenching of SF and the morphological transformation of gas-rich, late-type galaxies into gas-poor early-type systems. The debate is not yet settled, however, concerning to what extent the formation (nature) or subsequent evolution of galaxies and the interaction with their surroundings (nurture) give rise to the observed morphology-density relation. Both aspects are likely to play a role but the relative importance of nature and nurture in the formation and evolution of galaxies in dense environments seems to depend on both cosmic time and environment.\\

Several gravitational and hydrodynamical mechanisms have been proposed as being responsible for the evolution of galaxies in dense environments such as galaxy groups and clusters \citep[e.g.][]{Gunn1972,Nulsen1982,Tinsley1979}. While gravitational interactions affect both the stellar and the gaseous components of galaxies, hydrodynamical interactions only influence the gaseous component. \citet{Gunn1972} made the first suggestion with their proposed hydrodynamical mechanism of ram-pressure stripping that the interaction between a galaxy's ISM and the intracluster medium might affect the evolution of a galaxy. Similarly, other mechanisms were proposed such as viscous stripping \citep{Nulsen1982}, starvation \citep{Larson1980}, and thermal evaporation \citep{Cowie1977} considering the effects of viscosity, turbulence and thermal conduction on the depletion of a galaxy's gas reservoir. Gravitational interactions like galaxy-galaxy tidal interactions and mergers \citep{Spitzer1951, Tinsley1979, Merritt1983, Springel2000}; galaxy-cluster interactions \citep{Byrd1990, Valluri1993}, and galaxy harassment \citep{Moore1996, Jaffe2016} mainly occur in denser environments such as galaxy groups and cluster outskirts. In the course of these interactions, the cold gas may also funnel towards the central regions of galaxies and act as fuel for nuclear SF or trigger an active galactic nucleus \citep[AGN;][]{Baldry2004, Poggianti2017, Deb2020}. Moreover, cluster galaxies may already have lost their gas and halted their SF activity prior to entering the dense cluster environment, termed as `pre-processing’ \citep{Solanes2001,  Tonnesen2007, Yoon2017}. It is not clear, however, which of these processes are dominant in which cosmic environment, and to what extent they impact the multi-phase ISM and the SF activity of the galaxies.\\

Since the gas in galaxies is the raw fuel for star formation \citep[e.g.][]{Larson1972, KennicuttJr.1998}, understanding galaxy evolution requires understanding the prevalence of the various physical processes that are responsible for the accretion and depletion of gas in galaxies in different environments. The extended, fragile, collisional \hi gas discs in galaxies serve as excellent diagnostic tracers of environment dependent gravitational and hydrodynamic processes. Observationally, it is well established that the atomic hydrogen discs in galaxies are affected in the dense cluster environments and that cluster galaxies have a smaller \hi gas reservoir and often disturbed \hi morphologies and kinematics compared to their field counterparts \citep[e.g.][]{Haynes1984a, Cayatte1990, Solanes2001, Gavazzi2005, Chung2009}.\\

Until recently, \hi studies of the environmental impact on galaxies in clusters have been limited to targeted surveys of galaxies based on their optical properties, typically late-type galaxies, for example, studies of the Virgo cluster \citep{Warmels1988, Cayatte1990, Cayatte1994, Chung2009}, Abell 262 \citep{Bravo-Alfaro1997}, the Coma cluster \citep{Bravo-Alfaro2000, Bravo-Alfaro2001}, Ursa Major \citep{Verheijen2001a}, and Abell 2670 \citep{Poggianti2001}. In some cases, volume-limited surveys provide a more complete picture such as studies of the Hydra cluster \citep{Mcmahon1993}, Hercules cluster \citep{Dickey1997}, Abell 2218 \citep{Zwaan2001}, Abell 496, Abell 754, Abell 85 \citep{Vangorkom2003}, Abell 370 \citep{Lah2007}, Abell 963 and Abell 2192 \citep{Verheijen2007, Gogate2020}, the ALFALFA Virgo cluster survey \citep{Kent2008}, the Antlia cluster \citep{Hess2015} and the Fornax cluster \citep{Loni2021}. These studies have provided some insight into the environmental effects on the \hi content of cluster galaxies. \hi imaging observations show that the \hi morphologies in the central cluster galaxies are truncated, asymmetric, or offset from the stellar discs, sometimes trailing the galaxy along its infall trajectory and providing evidence of ram pressure stripping (RPS) as observed in the Coma and Virgo clusters \citep[e.g.][]{Bravo-Alfaro2001, Chung2009}. For less massive clusters such as the Hydra cluster and Ursa Major, the \hi deficiencies in the central galaxies are less dramatic \citep{Mcmahon1993, Verheijen2001a, Loni2021}, signifying that those over-densities may be in an earlier stage of formation. Studies have shown \citep[e.g.][]{Solanes2001, Chung2009} that there are strong correlations between \hi deficiency, proximity to the cluster centre, and the high velocity with which a galaxy moves through the ICM of a cluster.\\

Wide-area, blind, \hi imaging surveys of galaxy clusters are indispensable to reliably distinguish between different environmental mechanisms. By combining the spatially resolved \hi morphologies and kinematics with information on the stellar component, star formation activity and the global and local environments of a galaxy, it becomes possible to distinguish whether hydrodynamical or gravitational processes dominate the gas depletion in cluster galaxies. With the first pilot observations for massive \hi imaging surveys with state-of-the-art telescopes such as MeerKAT \citep{Jonas2016} and the Australian SKA Pathfinder {\citep[ASKAP][]{Johnston2007,Johnston2008a}} telescope becoming available, it is timely to broaden our understanding of environmental processes acting on the ISM of galaxies in clusters and their outskirts.\\

In this paper we present \hi and radio continuum imaging data of the galaxy cluster Abell 2626 (hereafter A2626; \citealt{Abell1958}) obtained with the \mk telescope \citep{Jonas2016}. A2626 is a moderately massive ($\sim$\numunit{5 \times 10^{14}}{\msun}) galaxy cluster at a redshift of $z = 0.055$ \citep{Healy2020c}. From X-ray observations, A2626 is identified as a cooling-core cluster \citep{Rizza2000,Wong2008,Mcdonald2018}. Extensive X-ray \citep[e.g.]{Wong2008,Kadam2019}, optical \citep[e.g.][]{Fasano2005} and radio continuum \citep[e.g.][]{Rizza2000, Wong2008, Gitti2013, Ignesti2018, Ignesti2020} imaging is available for this cluster, as well as optical spectroscopy \citep[e.g.][]{Cava2009, Healy2020c} to confirm cluster membership. A2626 is located within the single-dish ALFA Arecibo Legacy Fast survey \citep[ALFALFA;][]{Haynes2011, Haynes2018} and a few cluster members have been detected in \hi by this survey.\\

To date, radio continuum studies at various frequencies have focused on the centre of the cluster where a peculiar, extended radio source known as the ``kite'' is located \citep[e.g.][]{Rizza2000, Wong2008, Gitti2013, Ignesti2018, Ignesti2020}. X-ray studies of the cluster \citep[e.g.][]{Rizza2000, Wong2008, Kadam2019}, as well as studies of the velocity distribution of galaxies within the cluster based on optical redshifts \citep{Mohr1996,Mohr1997} have shown evidence of a galaxy group merging with the core of the cluster. A recent, dedicated observing campaign with the multi-object {optical} spectrograph Hectospec on the {6.5-m} MMT facility, provided $\sim$1900 additional redshifts for galaxies with \numunit{r<20.4}{\text{mag}} out to $1.5\, \rvir$ \citep{Healy2020c}. {\citet{Healy2020c} performed a Dressler-Shectman analysis \citep{Dressler1988} which} resulted in the identification of two distinct kinematic substructures of galaxies well within the cluster virial radius (\rvir), as well as four distinct kinematic substructures on the outskirts of A2626, which are thought to be galaxy groups in the process of being accreted by the cluster. Thus, there is evidence of recent and ongoing accretion of galaxies and galaxy groups, making A2626 a particularly interesting laboratory to study multiple, ongoing, environment driven processes of gas depletion, star formation quenching and morphological transformations of infalling galaxies. We also note that A2626 is less massive than the well-studied Coma cluster but similar in mass to the nearby Virgo cluster while its dynamical state resembles that of the nearby and $\sim$10$\times$ less massive Fornax cluster \citep{Mohr1996, Paolillo2002}. Therefore, a comparison of the \hi properties of the galaxy populations in these four clusters may reveal whether the dynamical state or the mass of a cluster has an impact on the atomic gas content of its constituent galaxies. \\

Moreover, A2626 hosts six candidate ``jellyfish'' galaxies \citep{Poggianti2015}, which are potentially experiencing extreme ram-pressure stripping that is displacing the multi-phase gas into tails trailing the galaxy stellar body along its infall trajectory \citep[e.g.][]{Gullieuszik2018, Poggianti2019, Ramatsoku2019, Deb2020}. The confirmed, striking jellyfish galaxy JW100 has been at the centre of multi-wavelength (radio to X-ray) studies by the GASP (GAs Stripping Phenomena in galaxies, \citealt{Poggianti2017}) collaboration \citep[e.g.][]{Moretti2019, Poggianti2019}, aimed at understanding how the stripping process affects the stellar and multi-phase gas components of the galaxy. The \hi data presented in this paper will contribute to a better understanding of the jellyfish phenomenon by compelling case studies such as for JW100 {(Deb et al. in prep)}.\\

The principal goal, however, of the MeerKAT observations of A2626 presented in this work is to provide an \hi perspective on the various environment dependent astrophysical processes that govern the evolution of galaxies in a dense cluster and its surroundings, analysing the spatially resolved \hi discs of individual galaxies as well as employing the \hi stacking technique. \\

This paper is structured as follows: in \secref{sec:data}, we discuss the \mk observations. We describe the calibration and imaging process, as well as the tests carried out to evaluate the quality of the calibration in \secref{sec:mk_dataprocess}. In \secref{sec:sourcefinding}, we discuss how the source finding was performed on the \hi spectral line cube. In the subsequent Sections~\ref{sec:data_products}~and~\ref{sec:atlas}, we present the global properties of the detected \hi sources. Throughout this paper, we use \numunit{{\rm H}_0 = 70}{\kms} and $\Omega_{m} = 0.3$.

%--------------------------------------------------------------------

\section{MeerKAT Observations}
    \label{sec:data}

\renewcommand{\arraystretch}{1.1}
\begin{table}[!t]
    \centering
    \caption{\mk observations and imaging}
    \begin{tabular}{ll}
    \hline
     \noalign{\vspace{0.5mm}}
     \multicolumn{2}{l}{Observing parameters}                   \\
     \hline
     \noalign{\vspace{0.5mm}}
     Observation dates          & 15, 16, 17 July 2019          \\
     Flux/bandpass calibrator   & J1939-6342 \\
     Phase/gain calibrator      & J2253+1608 \\ 
     Pointing centre            & 23:36:31.00 +21:09:35.9   \\
     Total integration time     & \numunit{15}{\text{hr}} (\numunit{3\times 5}{\text{hr}}) \\
     Number of active dishes    & 61                    \\
     Shortest-longest baselines & \numunit{40-7709}{\text{m}}            \\
     Correlator mode            & 4k, dual polarisation     \\
     Primary beam FWHM          & $58'$ at \numunit{1420.406}{\mhz}       \\
     Total bandwidth            & \numunit{856}{\text{MHz}}  \\
     Frequency range            & 900--1670\,{\text{MHz}} \\
     Spectral resolution        & \numunit{208.966579}{\text{kHz}}  \\
     \hline\hline
     \noalign{\vspace{0.5mm}}
     \multicolumn{2}{l}{Continuum imaging properties}              \\
     \hline
     \noalign{\vspace{0.5mm}}
     Imaged area                & 2$\times$2 deg$^{\rm 2}$  \\
     Frequency ranges           & \numunit{960-1160}{\mhz}               \\
                                & \numunit{1300-1520}{\mhz}               \\
     Synthesised beam           & $10\arcsec.7 \times 12\arcsec.9$ (LF image, robust$ =-1$)          \\
                                & $8\arcsec.5 \times 13\arcsec.2$ (HF image, robust$ =-1$)          \\
     \hline\hline
     \noalign{\vspace{0.5mm}}
     \multicolumn{2}{l}{\hi imaging properties}              \\ 
     \hline
     \noalign{\vspace{0.5mm}}
     Imaged area                & 2$\times$2 deg$^{\rm 2}$  \\
     Frequency range            & \textbf{\numunit{1300-1430}{\mhz}}             \\
     Velocity coverage          & $-2011$ to $27 766\,\kms$     \\
     Velocity resolution        & \numunit{44.1 \times (1+z)} {\kms}          \\
     Synthesised beam           & $6\arcsec.2 \times 13\arcsec.7 $         \\
                                & (Robust$=-0.5$ at \numunit{1420}{\mhz})      \\
     Cube rms                   & \numunit{80}{\mu\text{Jy/beam}}  \\                                        
     \hline\hline
    \end{tabular}
    \label{tab:obsparam}
\end{table}

MeerKAT observations of A2626 were carried out over the nights of 15--17 July 2019, allocated in response to the first call for MeerKAT-64 open time proposals, which were on a shared risk basis (project code SCI-20190418-JH-01). For these first science observations, the correlator capacity was restricted to provide 4k channels across the entire band pass, resulting in a limited frequency resolution (see below). \\

The full width half maximum (FWHM) of the \mk primary beam provides a field-of-view (FoV) of $61\arcmin$ at 
a frequency of \numunit{1346}{\mhz} where \hi emission from $z\approx0.058$ is detected. This FWHM is adequate to encompass the \rvir radius of the cluster, which is \numunit{\rvir \approx 1.59}{\mpc}, corresponding to $25\arcmin$ at $z=0.058$ \citep{Healy2020c}. The sensitivity of \mk, however, enables the detection of \hi emission well beyond the FWHM of the primary beam and, therefore, we imaged an area of \numunit{2\times 2}{\text{deg}^2} centred on the cluster (see \secref{sec:hireduction} for details). Consequently, the \hi and radio continuum data presented in this publication may occasionally be corrected for a significant primary beam attenuation. \\

A2626 attains a maximum elevation of \numunit{38}{\deg} above the northern horizon of the \mk site, resulting in a rather elongated synthesised beam and necessitating relatively short tracks. A total integration time of \numunit{3 \times 5}{\text{hours}} was accumulated on-source for A2626. Scans of the flux/bandpass calibrator (J1939-6342) were made at the start and end of each track, while scans of the complex (amplitude and phase) gain calibrator (J2253+1608) were made every \numunit{20}{\min} for \numunit{2}{\min}. No separate polarisation calibrator was observed. The data were accumulated using the SKARAB correlator in full polarisation and 4k mode, with an integration time of \numunit{8}{\text{sec}} per visibility record. The total \mk bandwidth is \numunit{856}{\text{MHz}} with 4096 channels, which are \numunit{207}{\text{kHz}} wide corresponding to \numunit{45}{\kms} at $z=0$. In this work we only processed two subsets of the total bandwidth as follows: \numunit{960 - 1160}{\text{MHz}}, and \numunit{1300-1520}{\mhz} for the continuum imaging (see \secref{sec:continuum}); and \numunit{1300 - 1430}{\text{MHz}} for the \hi data (see \secref{sec:hireduction}). The velocity range covered by the \hi data cube spans from \numunit{cz=-2011}{\kms} to \numunit{cz=27766}{\kms}, which corresponds to a volume depth of \numunit{420}{\text{Mpc}}.

The large FoV of MeerKAT enabled us to observe the cluster using a single pointing. With \numunit{15}{\text{hours}} on source, the $3\sigma$ \hi mass detection limit for a galaxy with a line width of \numunit{300}{\kms} is \numunit{\mhi = 2\times 10^8}{\msun}, with a 5$\sigma$ column density sensitivity of \numunit{\mathrm{N}_\hi = 2.5 \times 10^{19}}{\text{cm}^{-2}} per channel when the data are smoothed to $30''$ resolution. The observational parameters can be found in \tabref{tab:obsparam}.
    
%--------------------------------------------------------------------
    
\begin{figure*}[ht]
    \centering
    \includegraphics[width=\linewidth]{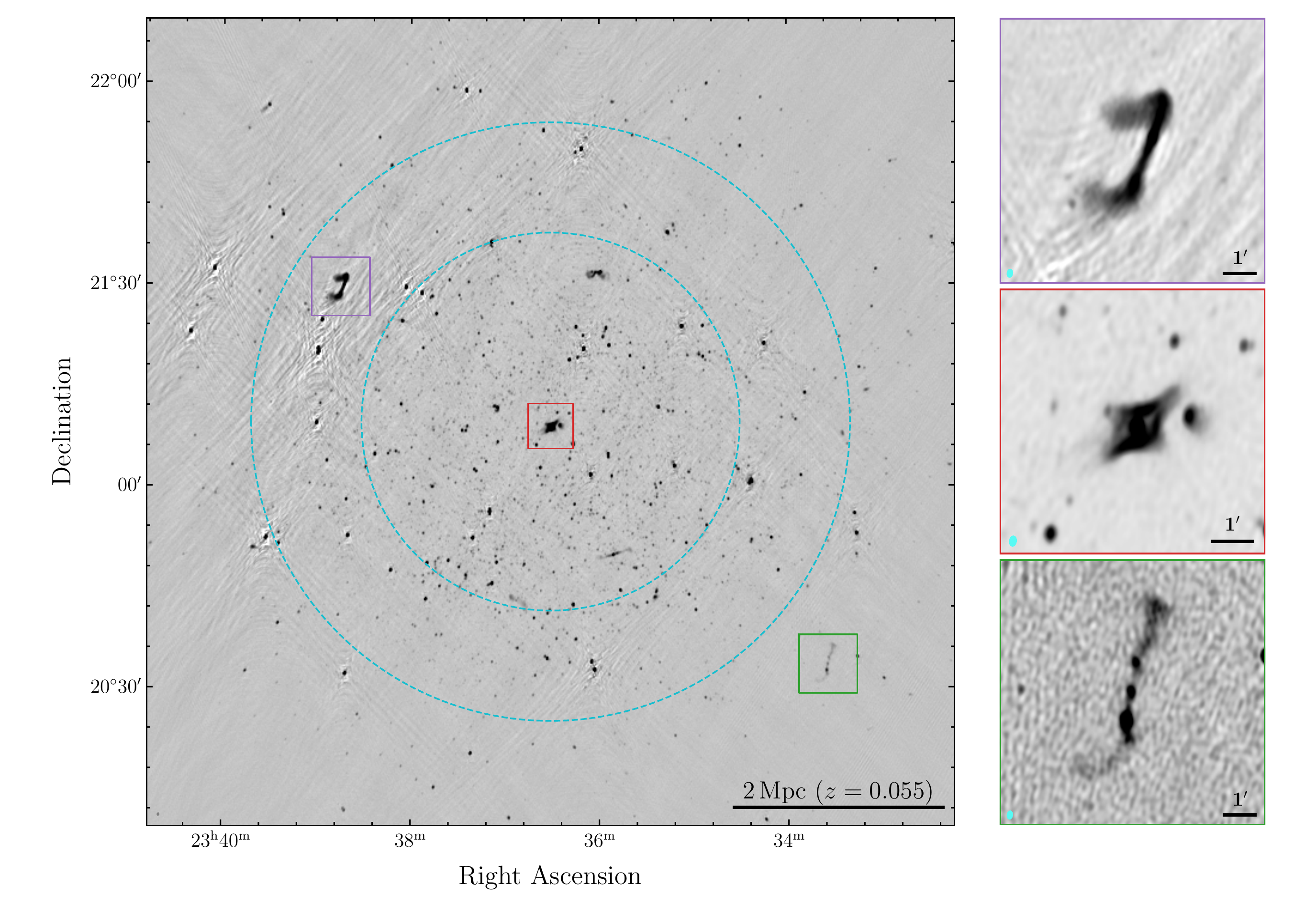}
    \caption{$2 \times 2 \,\mathrm{deg}^2$ L-band multi frequency synthesis continuum image centred on Abell 2626. This image covers a frequency range of 960--1160$\,$MHz and 1300--1520$\,$MHz, and was imaged with a robust weighting of -1. The image has not been primary beam corrected, however the FWHM of the primary beam at \numunit{1520}{\mhz} (the high frequency limit of the image) is given by the inner dashed cyan circle and the outer {dashed} cyan circle represents the primary beam at \numunit{960}{\mhz} (the low frequency limit of the image). The black bar in the bottom right corner indicates \numunit{2}{\text{Mpc}} at the distance of A2626 ($z = 0.055$). The synthesised beam ($8''.6 \times 12''.3$) is shown by the small grey ellipse in the bottom left corner. The three stacked panels on the right are zoomed in on two radio galaxies and the central ``kite'' source. The small filled cyan ellipse in the lower left corner indicates the synthesised beam.}
    \label{fig:continuum_image}
\end{figure*}

\section{Data processing}
    \label{sec:mk_dataprocess}

In order to limit the computing resources required to calibrate and Fourier transform the visibilities into images, we only processed a limited subset of the \numunit{856}{\mhz} total bandwidth. For the continuum image (see \secref{sec:continuum}) we processed two frequency ranges separately, \numunit{960 - 1160}{\mhz} and \numunit{1300 - 1520}{\mhz}, in order to avoid prolific Radio Frequency Interference (RFI) from navigation satellites. For the \hi data cube (see \secref{sec:hireduction} for details) we only processed the frequency range \numunit{1300 - 1430}{\mhz}. For both the continuum and \hi spectral line data, we only included the two orthogonal, horizontal (HH) and vertical (VV) polarisations in our data processing. The visibility data were calibrated on the ilifu computer cluster hosted by the Inter-University Institute for Data Intensive Astronomy (IDIA\footnote{www.idia.ac.za}) using the \caracal pipeline (version~1.0, formerly known as \texttt{MeerKATHI}, \citealt{Jozsa2020}). \\
   
We followed the same general strategy as \citet{Serra2019} for the cross-calibration using \caracal. Given the low altitude of A2626 at the \mk site, the data were flagged for geometric shadowing by nearby dishes. Possible RFI in the calibrator data was flagged using \texttt{AOFlagger} \citep{Offringa2012}, which is bundled with \caracal. We used the default `first pass' strategy of \texttt{AOFlagger} by only inspecting the Stokes Q amplitudes of the visibilities. The next step was to derive the antenna-based complex bandpass for the primary calibrator (J1939-6342). The bandpass solutions were bootstrapped to the secondary gain calibrator (J2253+1608) to determine its flux scale. The time dependent amplitude and phase corrections, as derived for each scan of the secondary gain calibrator, were applied to the target field by interpolating the gain corrections in time across the 20 minute time intervals. It is important to note here, that the gain solutions of the secondary calibrator were determined using only the baselines on which the calibrator was unresolved. This means that all projected baselines longer than \numunit{7}{\text{k}\lambda} were excluded as advised by the NRAO calibrator database. With all cross-calibration solutions applied to the target field, the target field was also flagged for RFI using the same 'first pass' \texttt{AOFlagger} strategy.

\subsection{L-band continuum imaging}
    \label{sec:continuum}
    
\begin{figure}[t]
    \centering
    \includegraphics[width=\linewidth]{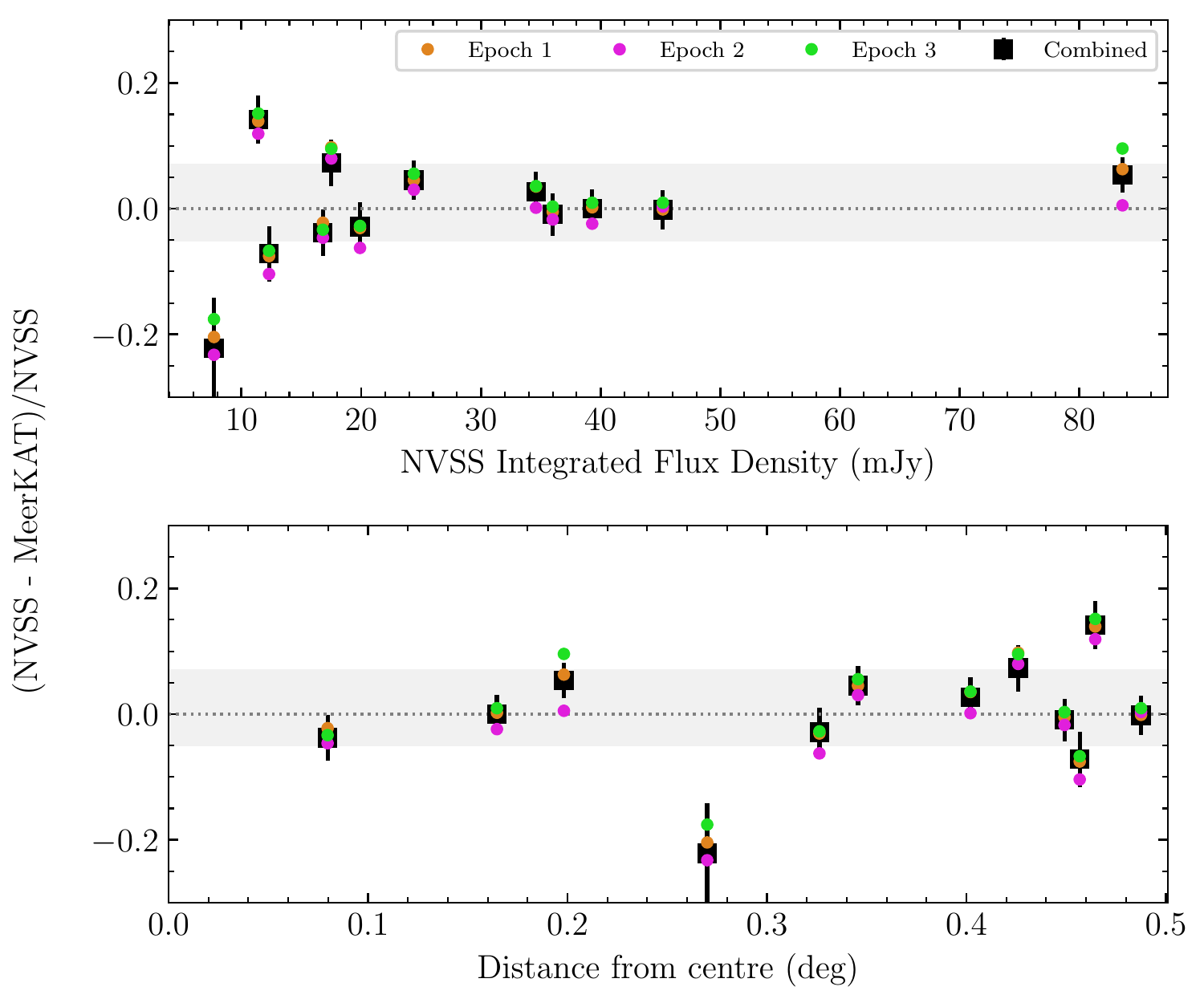}
    \caption{Comparison between the integrated fluxes from NVSS and MeerKAT (1300--1520$\,$MHz) for 12 sources unresolved by NVSS. Top: the fractional difference as a function of NVSS integrated flux density. Bottom: fractional difference as a function of radius from the centre of the pointing. The light grey horizontal band in each panel indicates the standard deviation of the differences.} 
    \label{fig:nvss_mk_comp}
\end{figure}

Using the cross-calibrated visibilities, we iteratively imaged and self-calibrated the target field using \texttt{WSClean} \citep{Offringa2014} and \texttt{CubiCal} \citep{Kenyon2018}. We performed three rounds of self-calibration with solution intervals of two minutes, performing two rounds of phase only, followed by one round of amplitude and phase. \\

We used multi-frequency synthesis (MFS) in \texttt{WSClean} to image the radio continuum of the target field for three frequency ranges: \textit{i}) \numunit{960-1160}{\text{MHz}} (LF image), \textit{ii}) \numunit{1300-1520}{\text{MHz}} (HF image), and \textit{iii}) \numunit{960-1160} combined with \numunit{1300-1520}{\text{MHz}} (full image). Shown in \figref{fig:continuum_image} is the \numunit{2\times 2}{\text{deg}^2} self-calibrated `full' MFS image constructed with a Briggs robust weighting of $r=-1$, giving an angular resolution of $8''.6 \times 12''.3$. The effect of primary beam attenuation can be readily appreciated as well as patches of limited dynamical range around bright sources near the half-power point of the primary beam. The three panels to the right of the main image in \figref{fig:continuum_image} show three extended radio galaxies, including the central ``kite'' source (middle panel), highlighting the resolution and surface brightness sensitivity of \mk. \\

To correct the images for primary beam attenuation, we calculated a model of the primary beam at the centre frequencies using \texttt{eidos} \citep{Asad2019}. The centre frequency used for the LF image is \numunit{1060}{\mhz} and for the HF image \numunit{1410}{\mhz}. For the full MFS image we used the average of the two primary beams calculated at \numunit{1160}{\mhz} and \numunit{1320}{\mhz}. We use an averaged primary beam rather than the \texttt{eidos} measured primary beam to approximate the primary beam at the centre frequency of \numunit{1240}{\mhz} for the full MFS image because we have effectively masked all data between \numunit{1160}{\mhz} and \numunit{1300}{\mhz}, and thus do not have data at the centre frequency. \\

We used \pybdsf \citep{Mohan2015} to identify continuum sources in the HF image, using the island threshold set to $2.5\sigma$ and the pixel threshold to $3.5\sigma$. We cross-matched our catalogue of continuum sources to a subset of unresolved sources in the NRAO VLA Sky Survey \citep[NVSS,][]{Condon1998}, using a search radius of $3''$. This cross-match yielded 12 sources within the FWHM of the \mk primary beam. Using the integrated flux density measurements from our \pybdsf catalogue, we compared the fractional difference between the NVSS flux densities and those from our \mk continuum images. We find good agreement between the NVSS and \mk integrated flux densities (see \figref{fig:nvss_mk_comp}) at all distances from the pointing centre and at all flux density values. {The offset and scatter between the two surveys is $0.9\%$ and $6\%$ respectively, this is shown by the grey bar in both panels of \figref{fig:nvss_mk_comp}}. In \figref{fig:nvss_mk_comp}, the errorbars on each point represent the uncertainty of the integrated flux density in the \mk image.

\subsection{H$\,$\textsc{i} line data reduction}
    \label{sec:hireduction}

\begin{figure*}
    \centering
    \includegraphics[width=\linewidth]{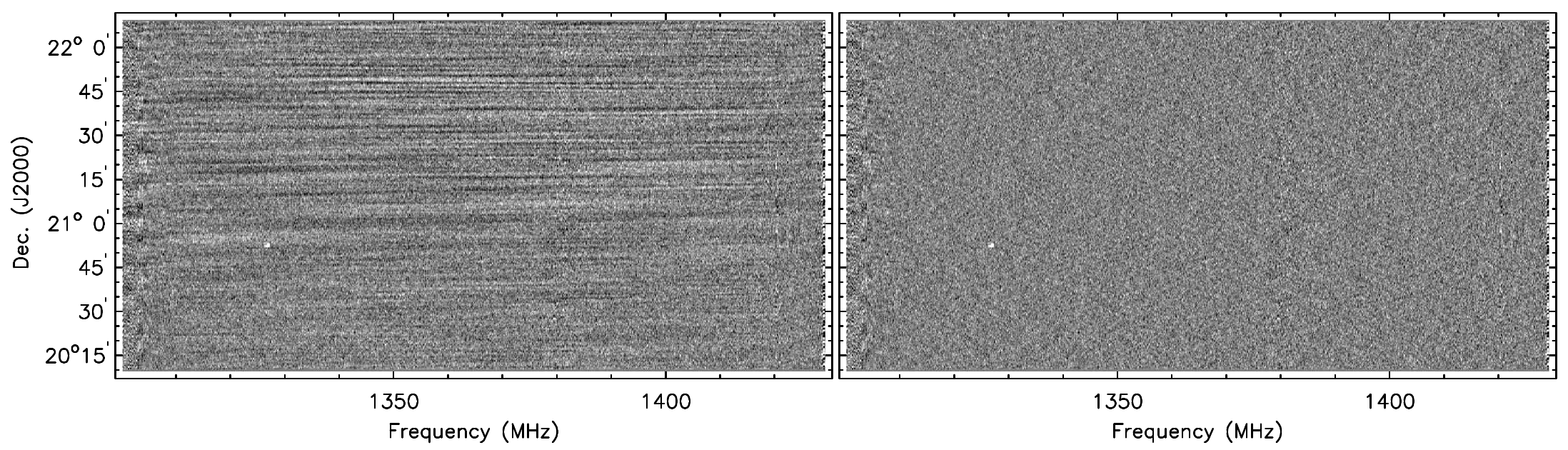}
    \caption{A vertical slice through the dirty data cube, smoothed to an angular resolution of $30''$, taken at R.A.=23:38:14.65 covering 2 degrees in declination and the full bandwidth. Left: Significant spectral ripples are clearly visible after subtracting the continuum model used for self-calibration, hampering faint source detection. Right: Illustration of the effectiveness of subtracting an iteratively fit spline function of order 15 from each spectrum, resulting in uniform noise. Note the \hi emission from a galaxy in the bottom-left quadrant of this position-velocity slice.}
    \label{fig:splinefit}
\end{figure*}

\begin{figure}
    \centering
    \includegraphics[width=\linewidth]{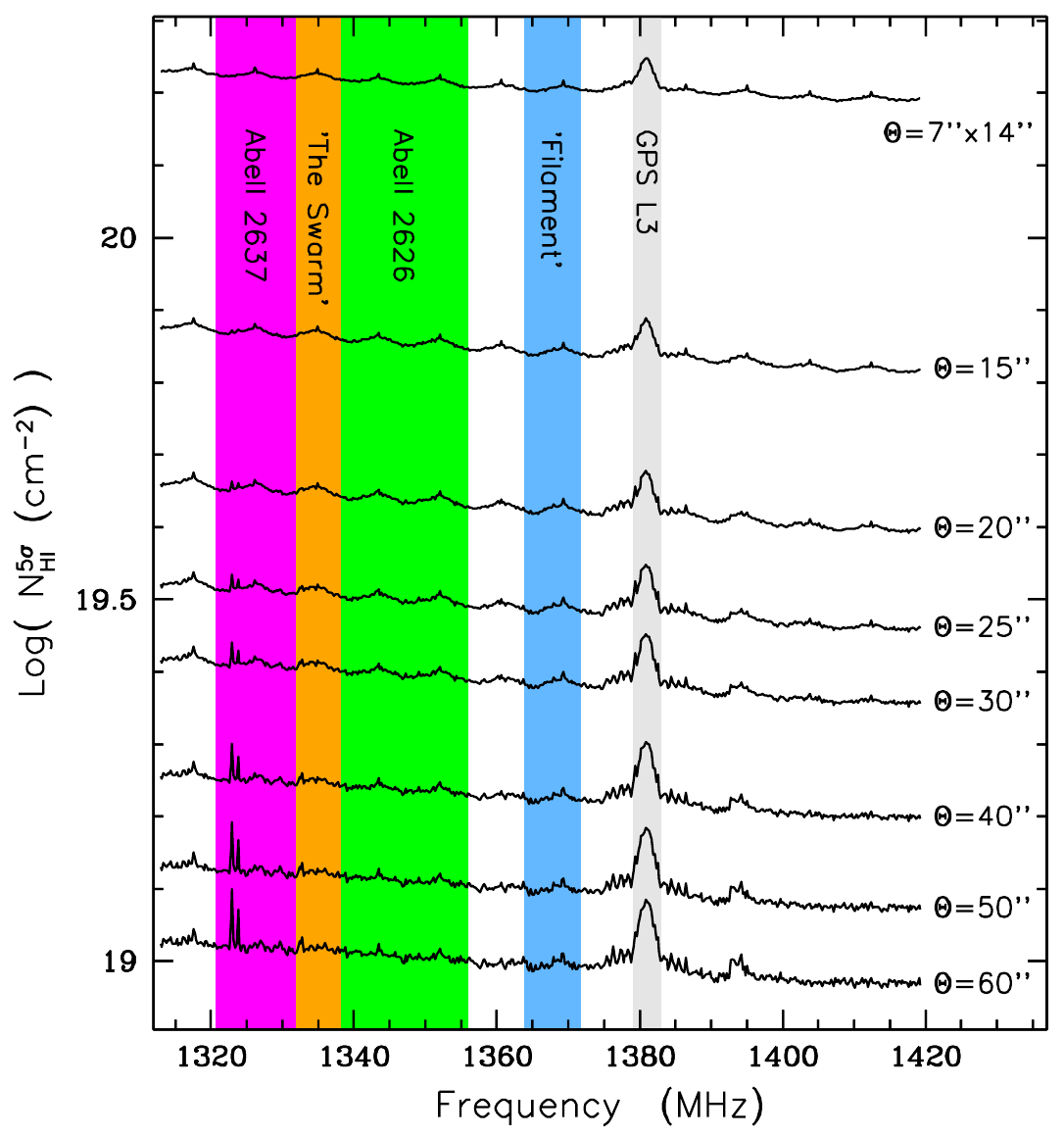}
    \caption{The 5$\sigma$ column density sensitivity per channel as a function of frequency and the various angular resolutions used in the source finding methodology. Vertical coloured bands indicate redshift ranges corresponding to cosmic over-densities as described in the text. {The grey band indicates the frequency of the GPS L3 band, an intermittent RFI source. }}
    \label{fig:noisefrequency}
\end{figure}

\begin{figure}
    \centering
    \includegraphics[width=\linewidth]{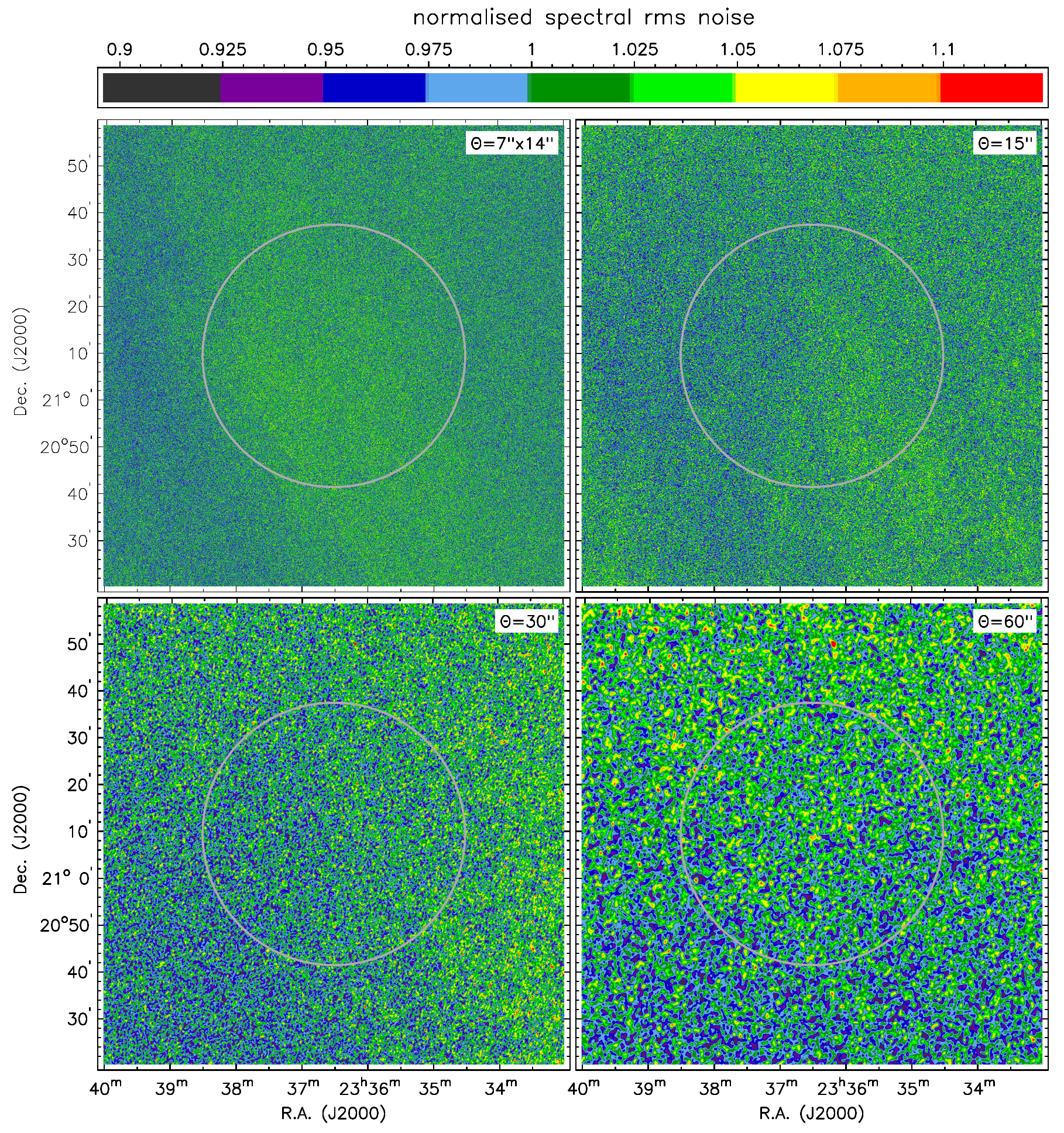}
    \caption{Normalised spectral noise maps at various angular resolutions, showing how the noise in the spectrum of each pixel varies across the field of view. The circle indicates the FWHM of the primary beam.}
    \label{fig:specnoise}
\end{figure}

We subtracted the sky-model created during the self-calibration process (see \secref{sec:continuum}) from the calibrated visibilities. Following the subtraction of the sky-model, we also subtracted a $4^{th}$ order polynomial fit to the visibilities to remove any residual continuum. The data were Doppler corrected to a barycentric reference frame and then imaged channel by channel using \texttt{WSClean} with a Briggs robust weighting $r = -0.5$, a pixel scale of $2''.5$, and no further UV-tapering. Our choice of Briggs weighting is motivated by an informed compromise between angular resolution, \hi column density sensitivity and the level of sidelobes of the synthesised beam. We imaged the data over a frequency range of \numunit{1300-1430}{\mhz}, which covers the cluster at \numunit{\sim 1346}{\mhz}, foreground galaxies (including Milky Way emission), and some cosmic background structures out to $z \sim 0.09$.\\

Before cleaning, {we noticed a ripple in the dirty \hi cube} along the frequency axis (see \figref{fig:splinefit}). Having investigated possible sources thoroughly, we have not been able to identify the origin of this baseline ripple, and it is not something that has been identified in other \mk \hi data. In order to remove this ripple, we fitted a $15^{th}$ order spline to the spectrum of each pixel. This was done over three iterations where the spline was fit and subtracted from the data, followed by the clipping of outlying pixels including the \hi emission from galaxies. Note that we chose the order of the spline to be sufficiently low to avoid fitting \hi emission from galaxies. {Note that, with a $15^{th}$ order spline, we remove spectral structures of order \numunit{8.7}{\mhz}, or $\sim$1850 \kms, which is much larger than the widest \hi emission line. We guarded against including \hi emission in the baseline fit by employing an iterative process, clipping any `emission' above and below the baseline subtracted spectrum and excluding those data from a next spline fit. Had this iterative process not worked satisfactorily, there would have been negative `bowls' on either side of the \hi emission and this is not what we observe.} The initial ripple is evident in the left panel of \figref{fig:splinefit}. The final ``de-rippled'' cube with a near uniform noise is shown in the right-hand panel of \figref{fig:splinefit}. \\

From the antenna pattern of the synthesised beam created during the imaging process, we could measure the angular resolution of the \hi cube prior to cleaning. The native resolution of the \hi cube varies with frequency: $14''.7$ to $13''.5$ in the north-south direction, and $6''.7$ to $6''.1$ in the east-west direction. \\

The deconvolution or \textit{cleaning} of the \hi emission in the ``de-rippled'' cube is not straight forward. Since we do not know a priori where the \hi emission in the cube will occur, the source finding, masking and \textit{cleaning} are intertwined and iterative processes that will be explained in the following section.

%--------------------------------------------------------------------

\section{\hi source-finding and cleaning}
    \label{sec:sourcefinding}

The art of source-finding in an \hi blind imaging survey is not a uniquely defined problem. The detection criteria depend on instrumental characteristics such as angular and velocity resolution, noise characteristics and uniformity, the presence of side lobes of the synthesised beam and possible residuals after continuum subtraction. The characteristics of the galaxy population and associated \hi properties also play a role in setting the optimum detection criteria, and the source-finding method of choice can be tailored according to the science motivation of a particular survey. It should also be noted that balancing the trade off between completeness and reliability of the detected source population is a major challenge that usually requires an empirical approach.\\

We have used \sofia \citep{Serra2015}, the {So}urce {Fi}nding {A}pplication, which is specifically designed for automatic, reliable \hi source-finding in interferometric spectral-line data. \sofia is publicly available at GitHub. We used the \sofia-2\footnote{see \url{https://github.com/SoFiA-Admin/SoFiA-2}} package (Westermeier et al., 2021) for our analysis and included the corresponding parameter file we used for our source-finding run in Appendix~\ref{sec:sofia_param} to allow our results to be reproduced. \sofia offers several methods to find sources and we have applied the `smooth+clip' algorithm, which smooths the cube spatially and spectrally with multiple, user-defined Gaussian and boxcar kernels respectively. With each smoothing operation the local rms noise is determined and a 3.5-$\sigma$ clip level is applied. After the smoothing operations, \sofia evaluates the dimensions of volumes of connected pixels, and estimates the reliability of the detection by statistically comparing the density of positive and negative detections in a three-dimensional parameter space. If no negative detections with similar characteristics are found in the same area of parameter space, then it assigns a high reliability to the detection. We have required a 90\% reliability threshold for a detection to be accepted. \sofia-2 returns three-dimensional binary masks that we subsequently used to define the regions of \hi emission that should be \textit{cleaned}. The final masks were also used to construct data products such as the global \hi profiles, total \hi maps and \hi velocity fields as described in the next section. \\

The first step was to take our continuum subtracted, `dirty' data cube at its highest angular resolution and to identify and \textit{clean} the brightest \hi sources in order to remove the side lobes of the synthesised beam that could affect the detection of fainter sources. For this purpose we applied \sofia-2 with seven Gaussian smoothing kernels that resulted in angular resolutions of 15\arcsec, 20\arcsec, 25\arcsec, 30\arcsec, 40\arcsec, 50\arcsec and 60\arcsec, and two spectral boxcar kernels of three and five channels wide ($\sim$135 and $\sim$225 $\rm km s^{-1}$). The larger angular smoothing kernels exceed the typical {\hi} sizes of the galaxies at the distance of A2626 {($\sim$ 10\arcsec-15\arcsec)} but enhance the column density sensitivity of the data. This first run of \sofia-2 resulted in 111 three-dimensional \textit{clean} masks throughout the entire cube, many of which contained multiple, blended sources due to the relatively large angular smoothing kernels and the sheer number of gas-rich galaxies in the data cube. Nevertheless, we used these large \sofia-generated masks to \textit{clean} the \hi emission in the `dirty' cube down to 0.3$\times$ the rms noise in a channel using the standard H\"ogbom {CLEAN} algorithm \citep{Hogbom1974} in the \gipsy software package (Groningen Image Processing SYstem, \citealt{VanderHulst1992}). We \textit{restored} the \textit{clean} components with a Gaussian beam of FWHM {7\arcsec.1$\times$ 14\arcsec.3}, which is the FWHM of the synthesized beam at the frequency that corresponds to the redshift of A2626. This resulted in a high resolution data cube in which most \hi sources were \textit{cleaned}.\\

In the next step, we used this preliminary \textit{cleaned} data cube to refine the source-finding process and to construct \textit{clean} masks that are optimised for the individual galaxies. Given the sensitivity, redshift depth and relatively poor spectral resolution of our MeerKAT data, combined with the diversity of the galaxy population in our survey volume, ranging from nearby dwarfs to distant early-type galaxies, and the possibility of detecting tidal \hi features, we concluded that with a single, particular set of \sofia parameters and smoothing kernels we could not identify all \hi sources in the cube. \\

For example, the highest available angular resolution ({7\arcsec.1$\times$ 14\arcsec.3}) is most sensitive to the smallest and unresolved galaxies, albeit with a by-product of yielding more false detections due to a larger number of resolution elements in the cube. However, at that higher resolution, some of the faint extended sources remain undetected due to the limited column-density sensitivity and hence do not exceed the 3.5-$\sigma$ detection threshold. By smoothing the cube to lower resolutions, (50\arcsec or 60\arcsec) we improve the column-density sensitivity, thus bringing the faint extended galaxies above the detection threshold as the extended sources better match the larger beam. However, at those large smoothing kernels, we dilute the signal of the smallest galaxies that were barely detected at the highest angular resolution. \figref{fig:noisefrequency} shows the 5$\sigma$ column density sensitivity as a function of frequency at the different angular resolutions.   \\

Therefore, at first we smoothed the original resolution cube to a circular beam of 15\arcsec\ using \gipsy to avoid a detection bias in favour of emission that is extended in the north-south direction. Subsequently, we smoothed the 15\arcsec\ \hi cube in \sofia-2 to different spatial and spectral resolutions and experimented with several detection thresholds and reliability settings in \sofia-2 in order to explore the trade off between the number of false positives and real sources. Specifically, we smoothed the 15\arcsec\ \hi cube to the resolutions mentioned in \tabref{tab:smoothingkernals}. The normalised spectral noise maps for the native {7\arcsec.1$\times$ 14\arcsec.3} resolution, as well as three circular resolutions ($15''$, $30''$, and $60''$) are presented in \figref{fig:specnoise} and show that the spectral noise is relatively constant across the entire imaged FoV, varying by no more than roughly $\pm$10\%.\\

\renewcommand{\arraystretch}{1.1}
\begin{table}
    \centering
    \caption{Resolutions at which the four \sofia-2 runs searched for \hi sources.}
    \begin{tabular}{cll}\hline
\sofia-2  &   Gaussian & boxcar                   \\ 
run &   arcsec & km/s \\\hline
 1 &    15  & 45, 135, 225\\
2 & 15, 20  & 45, 135, 225\\
3 & 15, 20, 25  & 45, 135, 225\\
4 & 15, 20, 25, 30 & 45, 135, 225\\
     \hline
    \end{tabular}
    \label{tab:smoothingkernals}
\end{table}

\begin{figure*}
    \centering
    \includegraphics[width=\linewidth]{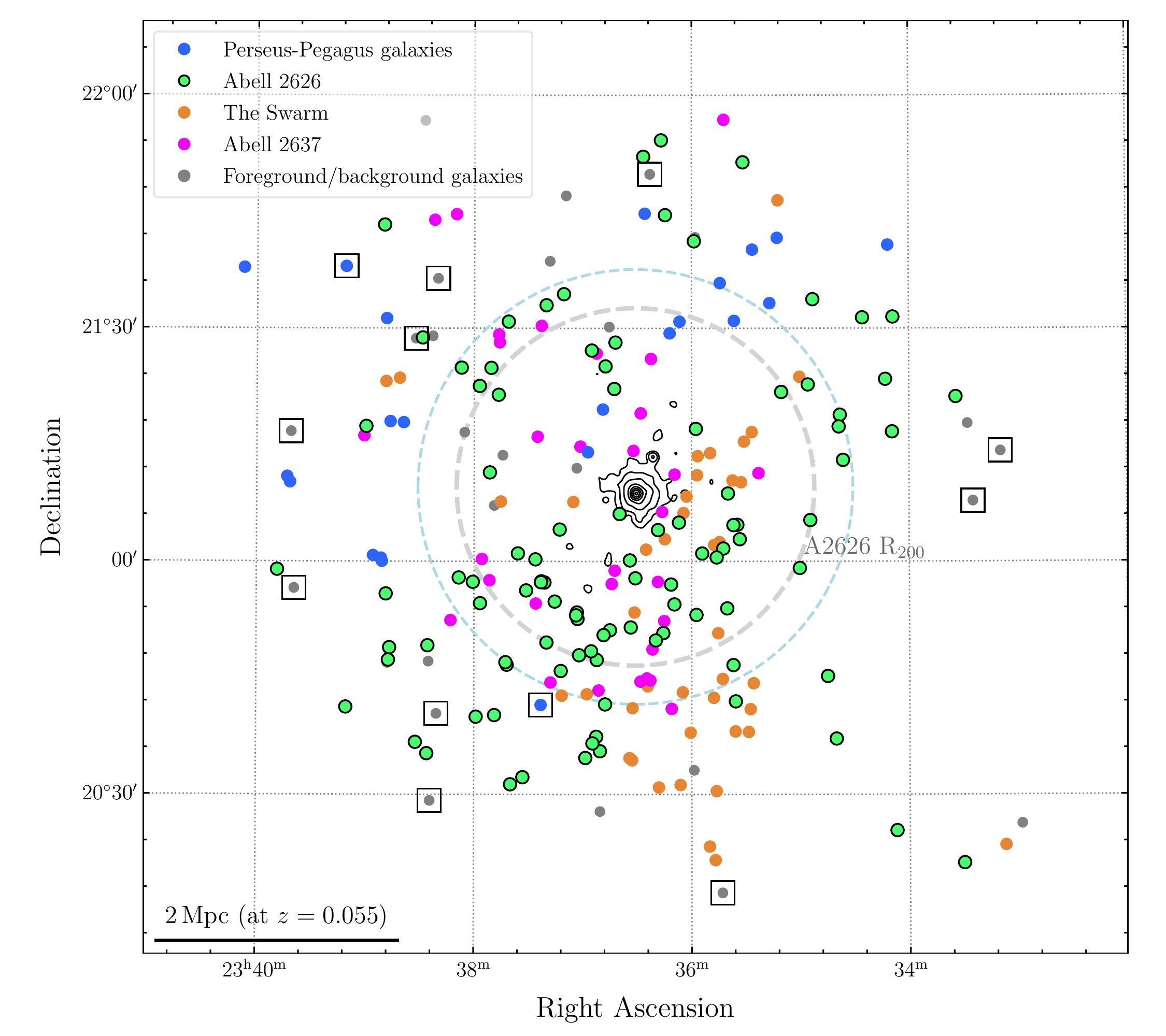}
  \caption{Sky distribution of the \hi detections identified with \sofia. The inner dashed grey circle represents the \rvir of A2626, and the outer light blue dashed circle represents the primary beam FWHM at the frequency of the cluster. The black contours represent the x-ray emission from A2626 taken from ROSAT images. The data points representing the \hi detections are coloured by the spectral over-density of which they are a member. {The blue indicates likely Perseus-Pegasus filament galaxies ($0.0475 < z < 0.0615$), green are A2626 galaxies ($0.0615 < z < 0.0675$), orange are galaxies in \hwone ($0.0675 < z < 0.0715$), and magenta galaxies represent galaxies part of A2637 ($0.0715 < z < 0.0745$); see \secref{sec:hi_dist} for more details.} The open squares represent the galaxies also detected by ALFALFA that are used in the \hi flux density comparison in \secref{sec:fluxcomp}.}
    \label{fig:a2626_skydist_hi}
\end{figure*}

Each of the four \sofia-2 runs has yielded a list of real and false positive detections, which may or may not be unique for these four different runs, primarily due to stochastic fluctuations. In a blind \hi detection experiment, most of the detections are actually very close to the detection threshold due to the slope of the \hi mass function. Hence, a small change in the source-finding parameters can result in the removal or addition of detections near the threshold. Thus, by merging the four runs we found some additional genuine detections that would otherwise have been lost due to stochastic fluctuations, thereby increasing the completeness of the catalogue but, at the same time, also accumulating false positives. The final merged list contained 404 unique objects, which may or may not be real \hi detections. \\

We assessed the reliability of each source by searching for a plausible {optical} counterpart within the \sofia-2 mask, taking advantage of images from the DECam Legacy Survey \citep[DECaLS][]{Dey2019}, Sloan Digital Sky Survey \citep[SDSS][]{York2000,Aguado2018} and Panoramic Survey Telescope and Rapid Response System \citep[Pan-STARRS][]{Chambers2016}. This resulted in 219 sources with a plausible optical counterpart and we {considered} the other 185 detections as false positives. {It is entirely possible, however, that some of these false positives may be actual sources of \hi emission associated with tidal debris or remnants of ram-pressure stripping events. Similarly, some of the \hi sources may have an optical counterpart as a chance superposition of a background galaxy for which no optical redshift is available.} For the 219 sources, {confirming} optical redshifts are available for 161 sources from the WIde-field Nearby Galaxy-cluster Survey \citep[WINGS][]{Cava2009}, SDSS \citep{Strauss2002,Aguado2018} and our MMT/Hectospec survey \citep{Healy2020c}. \\

The next task is to identify the optimum mask for each of the 219 \hi sources. If a source is detected in \sofia in only one of the four runs, then the \sofia mask generated by the corresponding run is considered as the optimum mask. In case an \hi source is detected in multiple runs, the mask from the highest run number is adopted as the optimum mask. Five \hi detections turned out to be blended sources, for which we manually de-blended the masks {to disentangle the different \hi sources in the masks produced by \sofia. De-blending was done for the following pairs of galaxies: 38, 42, 108, 109, 163, 164, 207, 208, 216, and 217.}
\\

The final step is to \textit{clean} all of the \hi sources since for the initial \textit{clean}ing we have used the masks from a different list of sources (111 sources as mentioned earlier) that does not contain all the 219 \hi sources of our final list. For the purpose of \textit{clean}ing, we have dilated the optimum masks for all 219 sources by 2 pixels (5\arcsec) outwards to include emission in the wings of the synthesised beam. We have subsequently used the dilated, optimum masks to define the regions of \hi emission in the highest resolution \hi cube and \textit{clean}ed down to 0.3$\sigma$. The \textit{clean} components are \textit{restored} with a Gaussian beam of similar FWHM as the dirty synthesised beam ($7\arcsec.1 \times 14\arcsec.3$), independent of frequency. Finally, we have smoothed the \textit{clean}ed high resolution cube to a circular beam of 15$\arcsec$ for further analysis.\\

\figref{fig:a2626_skydist_hi} shows the spatial distribution of the 219 \hi detections. The colours of the sources are based {to which over-density they belong}. Galaxies belonging to A2626 are coloured green, galaxies associated with \hwone are orange, and the pink points represent galaxies that are associated with Abell 2637. The blue symbols represent galaxies that have the same redshift as the Perseus-Pegasus filament at this location (see Fig.~8 in \citealt{Healy2020c}).

%--------------------------------------------------------------------
    
\section{\hi data products}
    \label{sec:data_products}

In this section, we describe the methods we adopted to derive the \hi data products and how we determined the \hi properties of our sample of 219 \hi detected galaxies {with optical counterparts}.

\begin{table*}[!t]
\renewcommand{\arraystretch}{1.1}
\setlength{\tabcolsep}{8pt} 
\centering
\caption{\textbf{A comparison of the central velocities, line widths and integrated fluxes} between the \mk survey presented in this work, and values from the ALFALFA survey.} 
\centerline{
\footnotesize{
\begin{tabular}{l|rrrrrrrl|rrrrl}\hline
\noalign{\vspace{0.5mm}}
 & \multicolumn{8}{|c}{This study} & \multicolumn{5}{|c}{ALFALFA} \\
\multicolumn{1}{c}{Name}      &
\multicolumn{1}{|c}{$cz$}  & 
\multicolumn{1}{c}{$\pm$}  & 
\multicolumn{1}{c}{\wtwenty}  & 
\multicolumn{1}{c}{$\pm$}     &
\multicolumn{1}{c}{W$_{50}$}  &
\multicolumn{1}{c}{$\pm$}     &
\multicolumn{1}{c}{$\int$Sdv}    & 
\multicolumn{1}{c}{$\pm$}     &
\multicolumn{1}{|c}{$cz$}     & 
\multicolumn{1}{c}{W$_{20}$}     & 
\multicolumn{1}{c}{W$_{50}$}     & 
\multicolumn{1}{c}{$\int$Sdv} &
\multicolumn{1}{c}{$\pm$}  \\

 & \multicolumn{2}{|c}{(\kms)} &
\multicolumn{2}{c}{(\kms)} &
\multicolumn{2}{c}{(\kms)} &
\multicolumn{2}{c}{(Jy$\,$\kms)} &
\multicolumn{1}{|c}{(\kms)} &
\multicolumn{1}{c}{(\kms)} &
\multicolumn{1}{c}{(\kms)} &
\multicolumn{2}{c}{(Jy$\,$\kms)}  \\

\multicolumn{1}{c}{(1)} &
\multicolumn{1}{|c}{(2)} &
\multicolumn{1}{c}{(3)} &
\multicolumn{1}{c}{(4)} &
\multicolumn{1}{c}{(5)} &
\multicolumn{1}{c}{(6)} &
\multicolumn{1}{c}{(7)} &
\multicolumn{1}{c}{(8)} &
\multicolumn{1}{c}{(9)} &
\multicolumn{1}{|c}{(10)} &
\multicolumn{1}{c}{(11)} &
\multicolumn{1}{c}{(12)} &
\multicolumn{1}{c}{(13)} &
\multicolumn{1}{c}{(14)} \\
\hline
\hline
\noalign{\vspace{0.5mm}}
J233309.58+211415.0 &  5790 &  3 & 388 &  3 & 358 &  4 & 12.83 & 0.24 &  5789 & 384 & 352 & 11.88 & 0.13 \\ 
J233324.80+210748.4 &  5725 &  7 & 437 &  7 & 401 &  9 &  5.78 & 0.43 &  5722 & 430 & 396 &  6.47 & 0.14 \\ 
J233543.02+201720.8 &  2187 & 20 & 260 & 20 & 228 &  7 &  5.60 & 0.15 &  2173 & 287 & 242 &  5.25 & 0.09 \\ 
J233623.11+214949.6 &  9447 & 27 & 234 & 28 & 186 & 16 &  0.94 & 0.06 &  9436 & 231 & 181 &  0.97 & 0.08 \\ 
J233723.19+204131.7 & 11751 & 12 & 266 & 12 & 235 & 10 &  0.66 & 0.04 & 11745 & 256 & 222 &  1.02 & 0.08 \\ 
J233820.06+213626.4 &  9766 & 12 & 248 & 12 & 211 & 12 &  1.06 & 0.05 &  9774 & 212 & 184 &  1.13 & 0.09 \\ 
J233820.77+204025.3 & 13323 & 17 & 173 & 18 & 137 & 16 &  0.58 & 0.05 & 13316 & 205 & 158 &  1.13 & 0.09 \\ 
J233824.36+202913.1 &  2183 & 22 & 144 & 22 &  92 & 17 &  1.40 & 0.13 &  2183 & 126 &  73 &  1.39 & 0.06 \\ 
J233832.19+212841.9 &  5633 & 18 & 137 & 18 &  70 & 20 &  0.53 & 0.06 &  5634 & 146 &  79 &  1.00 & 0.06 \\ 
J233910.80+213758.6 & 11643 & 41 & 316 & 42 & 269 & 35 &  0.85 & 0.11 & 11647 & 253 & 210 &  1.17 & 0.08 \\ 
J233939.15+205633.4 &  4317 & 15 & 189 & 15 & 147 & 17 &  1.81 & 0.12 &  4329 & 198 & 150 &  1.95 & 0.08 \\ 
J233941.07+211643.9 &  2612 & 22 & 122 & 23 &  71 & 19 &  0.93 & 0.09 &  2577 & 120 &  62 &  0.70 & 0.04 \\ 
\hline
\hline
\end{tabular}
}
}
\label{tab:alfalfa}
\end{table*}

\subsection{\hi global profiles}
  \label{sec:higlobalprofiles}

The \hi global profiles were constructed by measuring the flux density within the spatially dilated \sofia masks on a channel-by-channel basis. These are the same masks that were used for the purpose of \textit{clean}-ing the \hi emission. For each galaxy, these 3-dimensional masks vary in shape, size and position from channel to channel due to the rotation of the \hi disk. In order to include possible missed emission from the extreme velocity edges of the \hi global profiles, for each galaxy we replicated the mask of the first channel to the preceding channel, and the mask of the last channel to the following channel. We used these masks to isolate the data pixels in each channel of the $15''$ resolution datacube and applied a channel dependent correction for primary beam attenuation based on \citet{Asad2019}. Subsequently, we used the \texttt{GIPSY} task \texttt{FLUX} to measure the spatially integrated flux density of the primary beam corrected data pixels within the masked region of each channel.\\

Since the size and shape of the masks are different across the frequency channels, the uncertainty in the measured global profile flux at each frequency channel also varies. For calculating the channel dependent uncertainties in the global profile flux, we have taken an empirical approach by projecting the \hi mask for a particular  channel of a galaxy at 24 different line-free, nearby locations in that same channel. After applying the primary beam correction, the flux inside each offset mask was measured and we calculated the variance of the 24 measurements as the uncertainty on the flux from the galaxy for that particular frequency channel.\\

We measured the observed widths of the global \hi profiles at the 20\% and 50\% levels of the peak flux. For Gaussian and asymmetric profiles we adopted the single, highest flux of the global profile. For clearly double-horned profiles, however, we adopted separate peak flux levels for the high and low frequency sides of the profile. The frequencies of the two data points above and below the 20\% and 50\% flux levels at both sides of the profile were linearly interpolated to determine the frequencies $\nu_{20}^{high}$, $\nu_{50}^{high}$, $\nu_{50}^{low}$ and $\nu_{20}^{low}$ on the high and low frequency sides of the global \hi profile. We note that these frequencies were determined `outside-in', which implies that we may have overestimated the width of the \hi profile in case the flux of the profile edges is not declining monotonically. The uncertainties on these frequencies were determined by considering the uncertainties in the channel-dependent flux levels.\\

The observed central frequency of the global profile is calculated according to:
\begin{equation}
    \nu^{obs} = 0.5 (\nu_{20}^{high} + \nu_{20}^{low})
\end{equation}

\noindent This central frequency is indicated with a black downward arrow in the global profiles as presented in the atlas pages (see Fig. \ref{fig:45_atlas} or Fig. \ref{fig:128_atlas} for example). From this observed frequency the galaxy's systemic recession velocity is calculated as
\begin{equation}
    {\rm V}_{sys} = cz_{\hi} = c(\;(\nu_0/\nu^{obs})-1)
\end{equation}

\noindent where $\nu_0$ is the rest frequency of the \hi emission line. The frequency widths of the global profiles are measured according to:
\begin{equation}
\Delta\nu_{20} = \nu_{20}^{high} - \nu_{20}^{low} \quad{\rm and}\quad \Delta\nu_{50}= \nu_{50}^{high} - \nu_{50}^{low}
\end{equation}

\noindent These frequency widths are indicated by grey horizontal double arrows in the global profiles as presented in the atlas pages (see Fig. \ref{fig:45_atlas} or Fig. \ref{fig:128_atlas} for example). From this, the rest frame observed velocity widths at the 20\% and 50\% levels were calculated according to:
\begin{equation}
 {\rm W}_{20}^{obs} = c(\Delta\nu_{20}/\nu^{obs}) \quad{\rm and}\quad {\rm W}_{50}^{obs} = c(\Delta\nu_{50}/\nu^{obs})
\end{equation}

\noindent The calculated values of V$_{sys}$,  W$_{20}^{obs}$ and W$_{50}^{obs}$ are tabulated in Appendix A along with their uncertainties based on the uncertainties in $\nu_{20,50}^{high,low}$. It should be noted that the reported line widths are measured as is and not corrected for the poor instrumental velocity resolution.\\

We integrated the global \hi profiles to obtain the integrated \hi flux densities $\rm S_V=\int S_\nu dV$ in units of Jy \kms and tabulated these values in Appendix A. From these integrated flux density values we infer the total \hi gas masses in units of $\rm M_\odot$ according to:
\begin{equation}
\rm \mhi =\frac{2.36 \times 10^5 D_{L}^2}{(1+z)} \int S_\nu dV
\end{equation}

\noindent where $\rm \int S_\nu dV$ is the integrated \hi flux density in Jy $\rm km s^{-1}$ and $\rm D_{L}$ is the luminosity distance to the galaxy in units of Mpc. The luminosity distance for each galaxy is determined using the galaxy redshift. The inferred \hi masses are presented in \figref{fig:a2626_mhi_z}b where the \mhi uncertainties are determined from the flux uncertainties in each channel of the global profile.

\subsubsection{\hi flux and profile comparison}
    \label{sec:fluxcomp}

\begin{figure*}
    \centering
    \includegraphics[width=\linewidth]{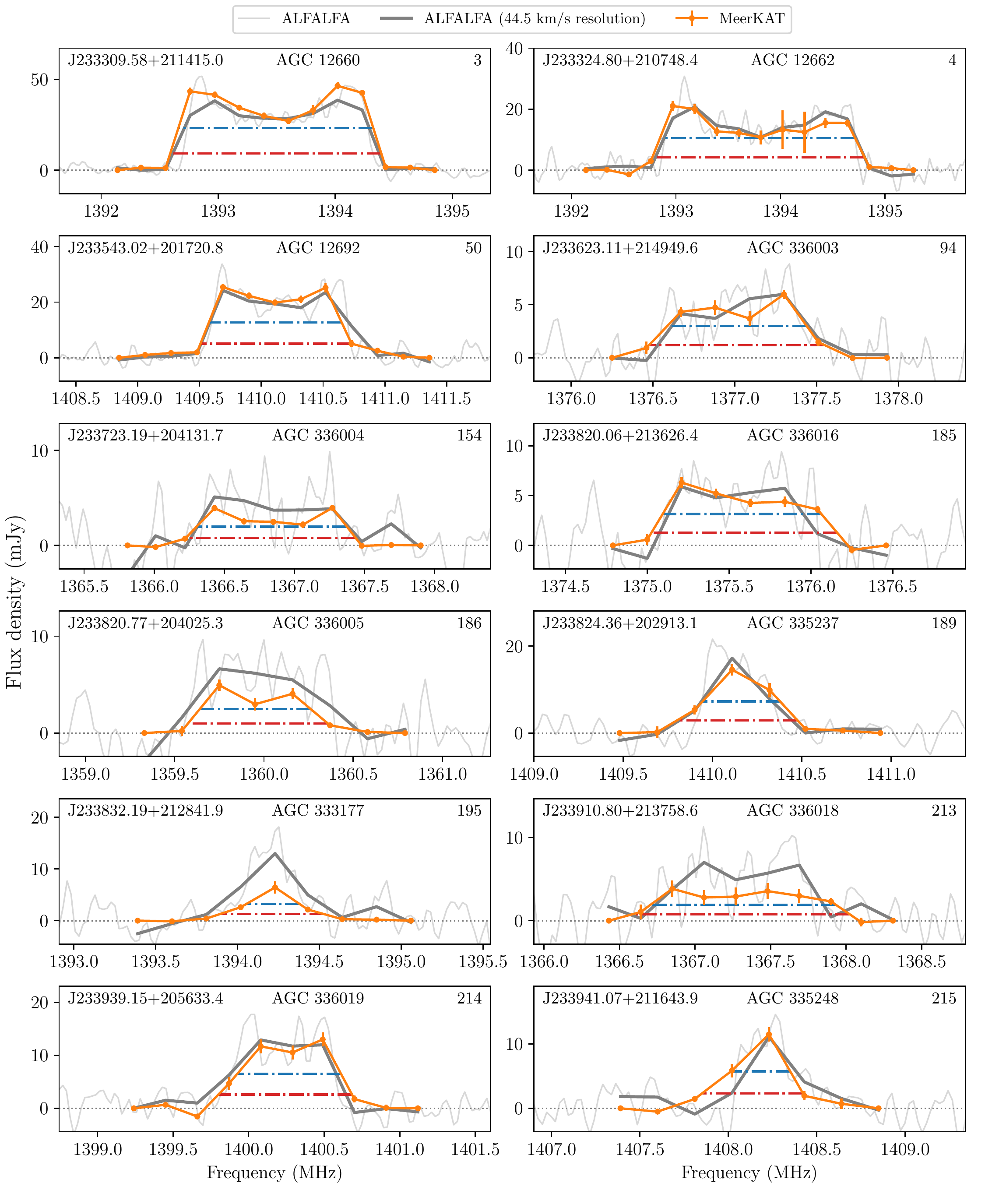}
    \caption{Global profiles for the 12 galaxies identified in the ALFALFA survey that are also identified in the \mk \hi data. The \mk spectra are shown in orange, the ALFALFA spectra in light grey (original spectral resolution) and dark grey (\mk spectral resolution). The red and blue horizontal dot-dash lines indicate where the \wtwenty and \wfifty respectively are measured for the \mk spectra.}
    \label{fig:a2626_alfalfa}
\end{figure*}

In the same way that we have compared the integrated flux densities of a number of sources in our continuum image (see \secref{sec:continuum}) to previously published flux densities of the same sources, we compare a sample of the integrated \hi flux densities to previously published values. In this case, we have an overlap of 12 sources with the Arecibo Legacy Fast ALFA survey \citep[ALFALFA;][]{Haynes2018}. The 12 sources have no known nearby galaxies that could possibly confuse the \hi spectrum{,
and are all classified as `code 1' detections in the ALFALFA catalogue (SNR$ > 6.5$)}. All 12 sources are located outside the FWHM of the \mk primary beam. The ALFALFA and \mk \hi spectra for the 12 galaxies are shown in \figref{fig:a2626_alfalfa}, and the measured line widths and integrated flux densities are presented in \tabref{tab:alfalfa}. \\

\begin{figure}
    \centering
    \includegraphics[width=\linewidth]{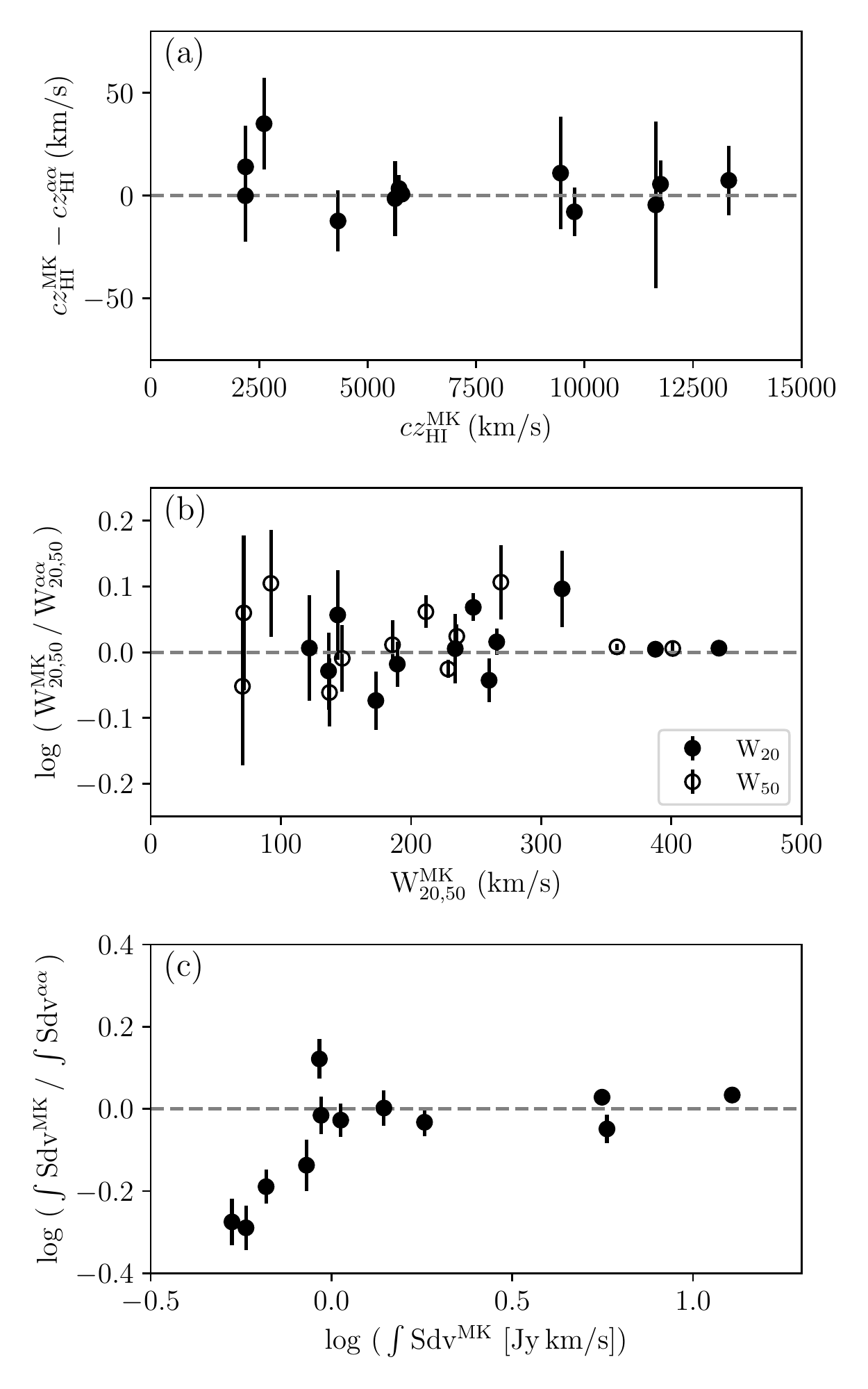}
    \caption{Comparison between measured properties from the global profiles of galaxies in both the \mk and ALFALFA surveys. (a) the difference between the recessional velocities calculated from the \hi redshifts. (b) the log ratio of the line widths, open circles represent the \wfifty and filled circles represent the \wtwenty. (c) log ratio of the integrated flux densities.}
    \label{fig:alfalfa_mk_comp}
\end{figure}

\figref{fig:alfalfa_mk_comp} shows the comparison between ALFALFA and \mk for the properties measured from the \hi global profiles (see \secref{sec:higlobalprofiles}). The \hi redshift and integrated flux density measurements for the ALFALFA galaxies are taken directly from the $\alpha.100$ catalogue \citep{Haynes2018} {as listed \tabref{tab:alfalfa}.}. Using the same procedure as discussed in \secref{sec:higlobalprofiles}, we re-measured the \wtwenty and \wfifty line widths from the ALFALFA spectra {(\numunit{R = 11}{\kms})} after they were re-sampled to the same spectral resolution as the \mk spectra {(\numunit{R = 45}{\kms})}. {The re-measured line widths are presented in \tabref{tab:alfalfa}, and are used in the comparison of line widths between the two surveys.} \\

Panel (a) of \figref{fig:alfalfa_mk_comp} shows the comparison between the measured \hi redshifts, shown here as the recession velocity $cz_\hi$. The offset between the \mk measurements and ALFALFA is \numunit{4}{\kms} with a scatter of \numunit{11}{\kms}. Given that both the offset and the scatter are much less than the velocity width of a single channel (\numunit{45}{\kms}), we conclude that the measured \hi redshifts for the two surveys are equivalent.\\

In the lower two panels of \figref{fig:alfalfa_mk_comp}, we present the log ratios of the measured profile line widths (b) and the integrated flux densities (c). The \wtwenty measurements are plotted as filled circles, and the \wfifty measurements as open circles. The offset for the \wtwenty and \wfifty measurements are 1.8 per cent (\numunit{0.008}{\text{dex}}) and 4.7 per cent (\numunit{0.02}{\text{dex}}) respectively, while the scatters are 9.6 per cent (\numunit{0.04}{\text{dex}}) and 12 per cent (\numunit{0.05}{\text{dex}}) respectively. For the smallest galaxies in the overlap (\numunit{\int \mathrm{Sdv^{MK}} < 1}{\text{mJy}\,\kms}), ALFALFA measures slightly higher integrated flux density than \mk. The offset and scatter between the two surveys for galaxies with \numunit{\int \mathrm{Sdv^{MK}} > 1}{\text{mJy}\,\kms} is 1.6 per cent (\numunit{0.007}{\text{dex}}) and 12 per cent (\numunit{0.05}{\text{dex}}).

\subsection{Total \hi maps}

The total \hi maps display the integrated \hi column densities of the \hi gas along the line of sight. These maps are constructed by applying the same dilated, optimum 3-dimensional masks used to derive the global \hi profiles, in order to isolate the primary beam corrected data pixels in the \textit{clean}-ed data cube at $15''$ resolution that contain the \hi signal. The pixels outside the masks are set to zero and the pixel values inside the mask are summed along the frequency axis to obtain the 2-dimensional total \hi maps. The resulting map is then multiplied by the rest frame channel width in \kms to set the units of the pixel values to {Jy \kms/beam}. The advantage of this method is that we avoid adding noise from pixels without \hi signal and thereby obtain a higher signal-to-noise for certain pixels in the \hi maps. A complication of this method is that the noise becomes non-uniform across the \hi map. Consequently, the 3$\sigma$ column density level in the \hi maps is not well defined and will be different for each galaxy in our sample. Next, we explain how the pseudo-3$\sigma$ contour level for the total \hi map is estimated.\\

To define the approximate 3$\sigma$ column density level in a total \hi map, we constructed a signal-to-noise map for each galaxy. We took an empirical approach similar to how we estimated the uncertainties in the global \hi profiles. We shifted the 3-dimensional mask to 24 nearby, line-free locations in the $15''$ data cube and followed the procedure above to construct 24 2-dimensional maps that contain only noise. Subsequently, for corresponding pixels in these emission-free maps we calculated the variance of the 24 values to create a representative noise map for each galaxy. The total \hi map was divided by this noise map to obtain a signal-to-noise map. From this signal-to-noise map we selected the pixels with $2.5 < (\tfrac{S}{N}) < 3.5$ and calculated the average value of the corresponding pixels in the total \hi map. We adopt this average value as the representative 3$\sigma$ \hi column density level, being the lowest contour in the total \hi map of each galaxy that is reliable. Note, however, that this pseudo-3$\sigma$ contour may run through areas in the total \hi map where the signal-to-noise value may deviate significantly from 3. The signal-to-noise maps and the pseudo-3$\sigma$ contour are included in the atlas pages presented below.\\

Subsequently, the pixel values in the total \hi map are converted from {Jy \kms/beam} to \hi column densities in units of cm$^{-2}$ using:
\begin{equation}
\rm N_{HI}=1.83 \times 10^{18} (1+z) \int T_b dV
\end{equation}

\noindent where $\rm N_{HI}$ is the \hi column density in cm$^{-2}$, $\rm dV$ is the rest frame velocity width in \kms over which the emission line is integrated at each pixel in the map and $\rm T_{b}$ is the brightness temperature in K, which is calculated according to:
\begin{equation}
 \rm T_{b}=\frac{605.7}{\Theta_{x} \Theta_{y}} S_{\nu} {\Bigg(\frac{\nu_{0}}{\nu}\Bigg)}^{2}
\end{equation}

\noindent where $\rm S_\nu$ is the flux density in units of mJy/beam, $\rm \nu_{0}$ and $\rm \nu$ are the rest frequency and the observed frequency of the \hi emission respectively, and $\rm \Theta_{x}$ and $\rm \Theta_{y}$ are the major and minor axes of the Gaussian synthesized beam in arcseconds.

\subsection{\hi velocity fields}

A galaxy's \hi velocity field displays the rest frame line-of-sight component of the velocity vector along which the \hi gas moves in the gravitational potential of the galaxy. Usually the systemic heliocentric recession velocity, $cz$, is added to this projected velocity, indicated by the iso-velocity contour along the kinematic minor axis of the galaxy's rotating \hi gas disk.\\

Various methods can be adopted to construct a 2-dimensional \hi velocity field, ranging from calculating an intensity weighted first moment of the \hi spectrum at each pixel position, fitting (multiple) Gaussians, a one-sided Gaussian or a Gauss-Hermite function in an attempt to counteract the effects of beam smearing. The method of choice typically depends on the resolution and signal-to-noise of the data. We decided to fit a single Gaussian to the \hi spectrum at each pixel position in the total \hi map and used initial estimates for the centroid and dispersion of the Gaussian fit from an intensity weighted first and second moment calculation of the \hi spectra censored by the 3-dimensional mask. The peak flux density of the censored spectrum was adopted as the initial estimate for the amplitude of the Gaussian. With these initial estimates we fit single Gaussians to the full, uncensored spectra but force the baseline to zero. Uncertainties on the Gauss parameters are calculated on the basis of the noise in a spectrum.\\

Gaussian fits are accepted only when satisfying the following conditions: 1) the amplitude is larger than 3 times the rms noise in that profile, 2) the uncertainty in the velocity centroid of the Gaussian is less 1/3 of the width of a velocity channel, 3) the velocity dispersion is within the range 15-100 km/s considering cases of edge-on galaxies suffering from significant beam-smearing. We have to keep in mind that the velocity resolution is comparatively poor (\numunit{45}{\kms}) and the \hi emission for most galaxies is limited to only a few velocity channels (3-5). Consequently, for many galaxies, the velocity fields are patchy with holes due to limited signal-to-noise and, therefore, a lack of acceptable fits.

\subsection{\hi position-velocity diagrams}
\label{sec: pv diagrams}

Position-velocity diagrams of inclined rotating \hi gas disks are generally constructed by extracting a two-dimensional slice from the data cube, centred on the dynamical centre of a galaxy and in the direction of the kinematic major axis.
Such position-velocity diagrams demonstrate the projected shape of a galaxy's rotation curve and may indicate the presence of any kinematic asymmetry of the rotating gas disc. Usually, the kinematic major axes and dynamical centres are calculated from the two-dimensional velocity fields of spatially well resolved galaxies. Since most galaxies in our sample are not well resolved and the \hi velocity fields are not of the highest quality due to the poor velocity resolution of our data, we have adapted a different strategy for determining the central position and position angle of the position-velocity diagrams.\\

PAs were initially determined by visual inspection of the optical image while the receding side with respect to the systemic recession velocity of the galaxy was determined from the \hi kinematics, either from the two-dimensional velocity field or from the three-dimensional cube if the former was poorly defined. For nearly face-on galaxies with a poorly defined optical PA, the PA of the receding side was decided solely from the \hi velocity field. In case of ambiguity of both the optical image and the \hi velocity field, the PA was based on the elongation of the \hi map. In few cases, the PA was chosen to cut through interesting \hi morphological features that were not aligned with the optical or kinematic major axes of the galaxy. Instead of using the dynamical centres derived from the velocity fields of the rotating disks, we have adopted the optical centres of the galaxies from the SDSS source name. The position-velocity diagrams presented in the atlas pages (see \secref{sec:atlas} and Appendix B) were extracted from the 15$\arcsec$ resolution cube.

%--------------------------------------------------------------------
    
\section{The \hi catalogue and \hi Atlas}
    \label{sec:atlas}
    
    {In the previous section, we described the methodology used to produce the \hi data products. In this section, we discuss how the \hi atlas pages for the 219 \hi detections with optical counterparts are presented. }

\subsection{The \hi catalogue}

The observed \hi parameters of all 219 galaxies detected in our \mk survey are listed in table \ref{tab:catalogue} in Appendix A. These properties are derived from the global \hi profiles as presented in the atlas pages. Information that can be derived from the two-dimensional \hi column density maps and velocity fields, such as the size and asymmetry of the \hi gas disc and its kinematic position angle, will be presented in a forthcoming paper. The column entries of table \ref{tab:catalogue} are as follows:

\noindent Column (1): The assigned \hi identifier based on the ascending Right Ascension (J2000.0) of their optical counterparts as provided by SDSS (DR14).

\noindent Column (2): The SDSS identifier for the optical counterpart of the \hi detected galaxies, based on their Right Ascension and Declination (J2000.0)

\noindent Columns (3), (4): Heliocentric, systemic recession velocity ($cz$) based on the \hi redshift measured as the midpoint of the 20\% line width of the \hi global profile.

\noindent Columns (5), (6), (7) \& (8): The observed rest frame line width of the \hi global profiles and their uncertainties in km s$^{-1}$, computed at the 20\% and 50\% level of the \hi peak flux respectively, measured at a velocity resolution of \numunit{45\, (1+z)\,\,}{\kms}. See \secref{sec:higlobalprofiles} for details.

\noindent Column (9) \& (10): The measured, primary beam corrected, integrated \hi flux densities and their uncertainties in Jy \kms. 

\noindent Column (11) \& (12): {The radio continuum flux density} for the \hi detected sources and their uncertainties based on the primary beam corrected continuum maps with robust weighting, $r=-0.5$. 

\subsection{The \hi atlas}
\begin{figure}[t]
    \centering
    \includegraphics[clip, trim=0cm 8cm 0.5cm 1cm, width=\linewidth]{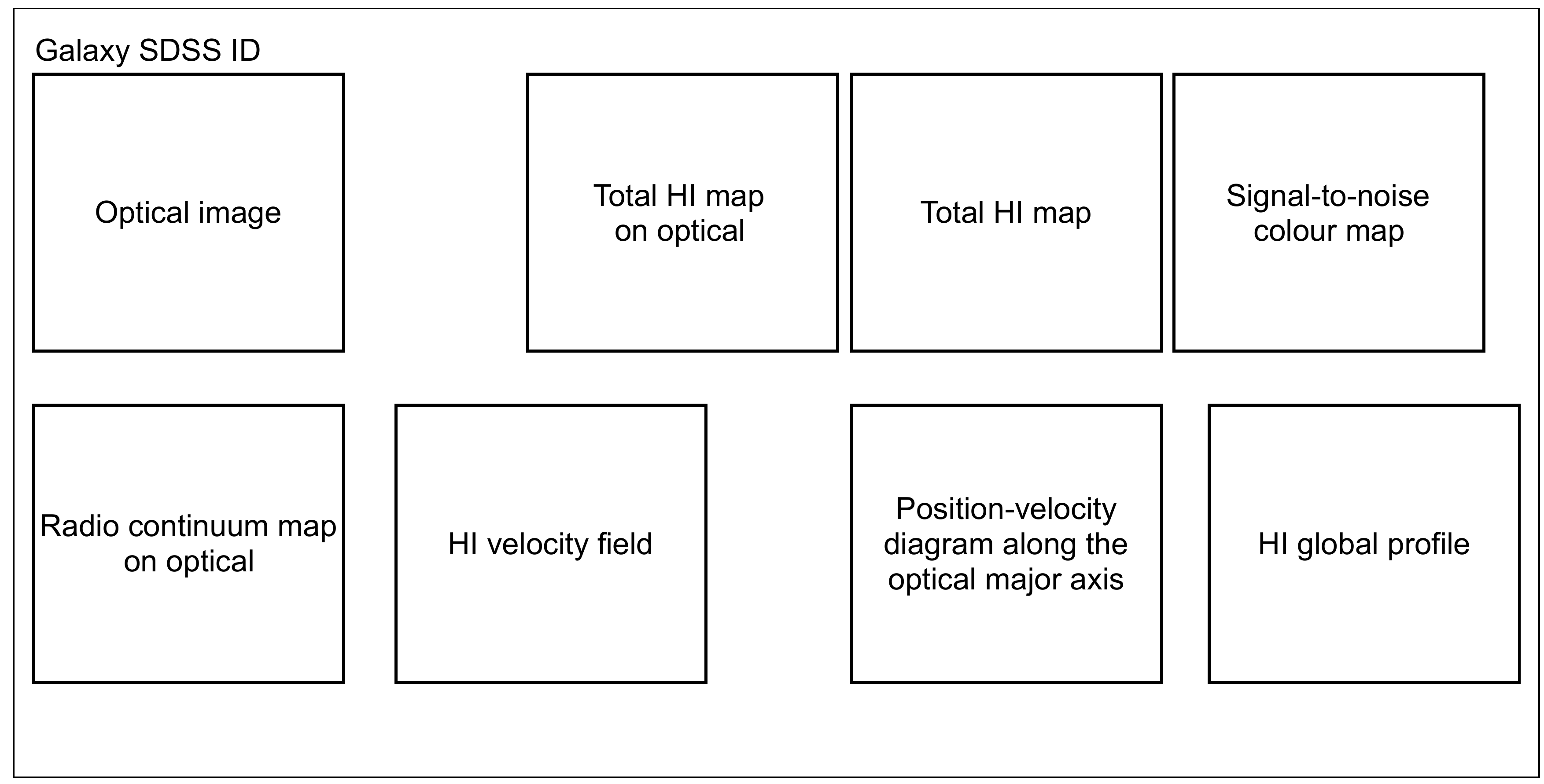}
    \caption{Layout of the \hi atlas for the 219 galaxies with fully reduced data. {See Appendix~\ref{app:atlas}.}} 
    \label{fig:atlas_page}
\end{figure}

The \hi data products as discussed in \secref{sec:data_products} are shown as atlas pages for all 219 \hi detected galaxies with fully reduced data. A schematic of the atlas pages is shown in \figref{fig:atlas_page}, each of the following subsections correspond to one of the panels on the atlas page.

\subsubsection{Optical image} A DECaLS r-band greyscale image is presented for each galaxy. The SDSS identifier of the galaxy associated with the \hi detection is shown above the image. The physical size of the image in terms of kpc at the \hi redshift of the galaxy is denoted by a scalebar in the bottom right of the panel. The orange open circle indicates the optical centre from the SDSS.

\subsubsection {\hi column density contours on optical image} {Contours of the \hi column density distribution are overlaid} on the DECaLS r-band image represented in greyscale in the background. For each galaxy, the \hi map is constructed based on the optimum \hi mask, which is the \sofia mask at the corresponding resolution of the \hi detection or the lowest resolution mask in case the \hi source was detected at multiple resolutions. The \hi column density contours are drawn at the levels N$^{3\sigma}_{\hi} \times 2^{\rm n}$ with n=0,1,2,3,...  while N$^{3\sigma}_{\hi}$, the average ${3\sigma}$ column density for each galaxy, is printed in the top-left corner of the signal-to-noise panel on the atlas page (see below). The FWHM of the beam size for each \hi source is indicated by the hatched circle in the bottom-left. The physical size of the map in units of kpc is presented as a scalebar in the bottom-right corner of the panel.

\subsubsection{Total \hi map} The total \hi map is presented in greyscale with the \hi column density contours superimposed. All positive pixels inside the total \hi map have an assigned greyscale value. The \hi column density contour levels are the same as the neighbouring aforementioned panel. Greyscale pixels outside the outermost contour typically have a signal-to-noise below 3. {The purple line indicates where we have extracted the position velocity diagram,} starting at the receding side of the galaxy (see Sec \ref{sec: pv_atlas}). The orange open circle signifies the optical centre from SDSS.

\subsubsection{Signal-to-noise map} The \hi signal-to-noise map is represented as a colour map with the different colours representing different levels of signal-to-noise as indicated by the vertical scale bar on the right of the panel. The average 3$\sigma$ \hi column density for each galaxy is calculated and printed in the top-left corner of the panel. The \hi contours are repeated in white, allowing to evaluate the signal-to-noise along the lowest contour and thereby its reliability.

\subsubsection{Radio continuum map on optical} The cleaned 21 cm radio continuum map is plotted as contours superimposed on the DECaLS r-band image in greyscale in the background. The continuum contours are drawn at the levels of $2^{\rm n} \times$ the rms noise, where n=0,1,2,3,.. . Dashed contours are drawn at $-2\times$ the rms noise. The rms noise levels are reported in the upper-left corner of the continuum map after correcting for primary beam attenuation and, therefore, may vary significantly for different sources. The FWHM of the beam size for each radio continuum source is indicated by the hatched circle in the bottom-left.

\subsubsection{\hi velocity field} {The greyscale image indicates where radial velocities could be measured.} Lighter greyscale and blue {iso-velocity} contours signify the approaching side of the galaxy while darker greyscale and red {iso-velocity} contours denote the receding side of the \hi disc. {The white contour indicates the systemic velocity derived from the \hi global profile.} The contours are drawn at intervals of $\pm$n$\times 25\,\,\kms$, both for the approaching and receding sides. The orange open circle in the middle indicates the optical centre from the SDSS.

\subsubsection{Position-velocity (PV) diagram} \label{sec: pv_atlas} {The position-velocity diagram is made along the major axis of the optical image,} starting at the receding side of the galaxy. The coordinates of the optical centre are adopted from the SDSS. The strategies we adopted for determining the central positions and the PAs are mentioned in Sec. \ref{sec: pv diagrams}.

 The PA of each galaxy is printed in the top-left corner of the panel. The directions of the start and end points of the slices are indicated  {respectively} in the bottom left and right corners of each panel. 

For all the galaxies, the PV diagrams were extracted from the cleaned cube at 15" resolution and at the \numunit{45\,}{\kms} velocity resolution. The contour levels are (-2, 2, 3, 4.5, 6, 9, 12, 15, 20, 25) $\times$ rms noise. The orange contour outlines the dilated, optimum mask as automatically generated by \sofia and used for each galaxy as described in section \ref{sec:sourcefinding}. The vertical dashed line corresponds to the optical centre of the galaxy. The horizontal dashed line indicates the central frequency as derived from the global profile.

\subsubsection{\hi global profile} The \hi spectrum or global profile of each galaxy is constructed by applying the dilated, optimum \hi mask for each galaxy to the $15''$ resolution datacube as described in \secref{sec:higlobalprofiles}. The method of deriving the errorbars as shown on the global profile is also described in \secref{sec:higlobalprofiles}. The vertical black downward arrow in the middle shows the central frequency, corresponding to the systemic velocity, which is the midpoint of the 20\% velocity width (\wtwenty). The \hi redshift is calculated from this central frequency and reported in the top right of the panel. For the galaxies with measured optical redshifts, an orange upward arrow is indicating their optical redshifts.  The grey, horizontal double-arrows indicate the measured line widths at the 20\% and 50\% level of the peak intensity as tabulated in the Appendix A.

%--------------------------------------------------------------------

\section{Sample properties of the \hi detections}
    \label{sec:hi_dist}
    
\begin{figure*}
    \centering
    \includegraphics[width=\linewidth]{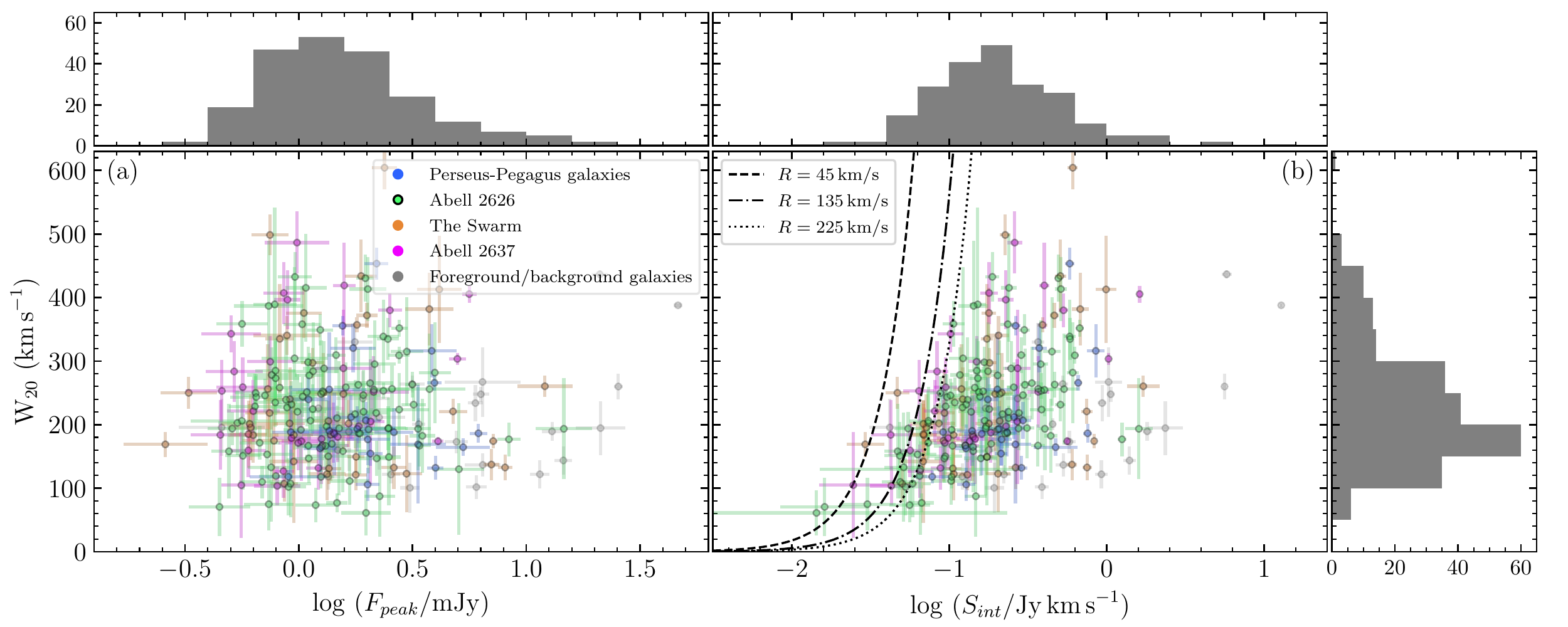}
    \caption{Bivariate distributions of measured properties from the global \hi profiles of the \hi detected sources. Peak flux {density (a) and integrated flux density (b) as a function of \wtwenty line width, derived from the global \hi profiles. The detection thresholds based on the three velocity smoothing kernels used for the source-finding are represented by the black lines in \figref{fig:a2626_distribution}b.} The colours of the data points represent the over-density to which the galaxy belongs: A2626 sources are green, \hwone sources are orange, A2637 sources are pink, and the sources coloured blue belong to the Perseus-Pegasus filament. Grey data points are galaxies in the volume that do not belong to a spectral over-density. The histograms along the top and right side of the panels show the 1D distribution of the observable along that axis. }
    \label{fig:a2626_distribution}
\end{figure*}

In this section we will briefly discuss the distributions of the measured parameters, the inferred \hi masses as a function of redshift and {the redshift distribution of galaxies} within the surveyed volume.

\subsection{Distributions of detection related parameters}

The detectability of a source depends on its distance (redshift) and \hi mass, resulting in an integrated \hi line flux density ($S_{int}=\int$Sdv), as well as its \hi line width (W$_{20}$), peak flux density ($F_{peak}$), angular size and distance from the centre of the primary beam, which attenuates the flux density from sources away from the pointing centre. In this subsection we explore the distributions of several of these observables \citep[e.g.][]{Kovac2009} to gain insight in the overall properties of our \hi detections.\\

\figref{fig:a2626_distribution} shows the \wtwenty line width of each galaxy as a function of the log of the peak flux density of the global profile (panel a) and the log of the integrated flux density of the galaxy (panel b).  The distributions show that there are no galaxies with large \wtwenty with small integrated flux densities and low peak flux densities in the observed volume. This is because galaxies with a small {integrated or peak flux densities} that is spread over a large velocity width would be very difficult to detect and might lie below our detection threshold. By smoothing in velocity and hence increasing the sensitivity, we could detect some galaxies with low flux densities and large line widths, {still our list of \hi sources is not complete.} The detection thresholds based on the three velocity smoothing kernels used for the source-finding are represented by the black lines in \figref{fig:a2626_distribution}b. The grey histograms atop panels a and b, and to the left of panel b in \figref{fig:a2626_distribution} show the one-dimensional distributions of $\log\,F_{peak}$, $\log\, S_{int}$, and \wtwenty respectively.\\

 The distribution of integrated flux density with redshift ($z$), presented in \figref{fig:a2626_mhi_z}a, shows a segregation of detections in the various over-densities {mentioned earlier: A2626, The Swarm and A2637.} We observe a lack of galaxies with {small $S_{int}$ (< -1.5 log S$_{int}$/Jy \kms) at low $z$ (z < 0.03 ).} This is likely due to two reasons, partly due to the smaller surveyed cosmic volume at low $z$, and partly since the low mass dwarf galaxies which we have the sensitivity to detect at this low redshift have velocity widths of the order of the channel width, thus escaping detection.
 In \figref{fig:a2626_mhi_z}a, lines of a constant \hi mass in \msun with a logarithmic step of \numunit{1}{\text{dex}} are shown as dotted lines. The \hi mass is calculated using Equation (5) where the distance is the luminosity distance calculated assuming \numunit{H_0 = 70}{\kms\,\mpc^{-1}} and $\Omega_m = 0.3$. We observe that most of the galaxies have \hi masses in the range of \numunit{10^{9}- 10^{10}}{\msun}. 
 At low redshifts, the cosmic volume we survey is much smaller and large galaxies are atypical. Furthermore, the large velocity channel width makes it very difficult to detect galaxies with narrow line widths (e.g. the low-mass dwarf population).

\subsection{\hi mass vs redshift}

 \begin{figure}
    \centering
  \includegraphics[width=\linewidth]{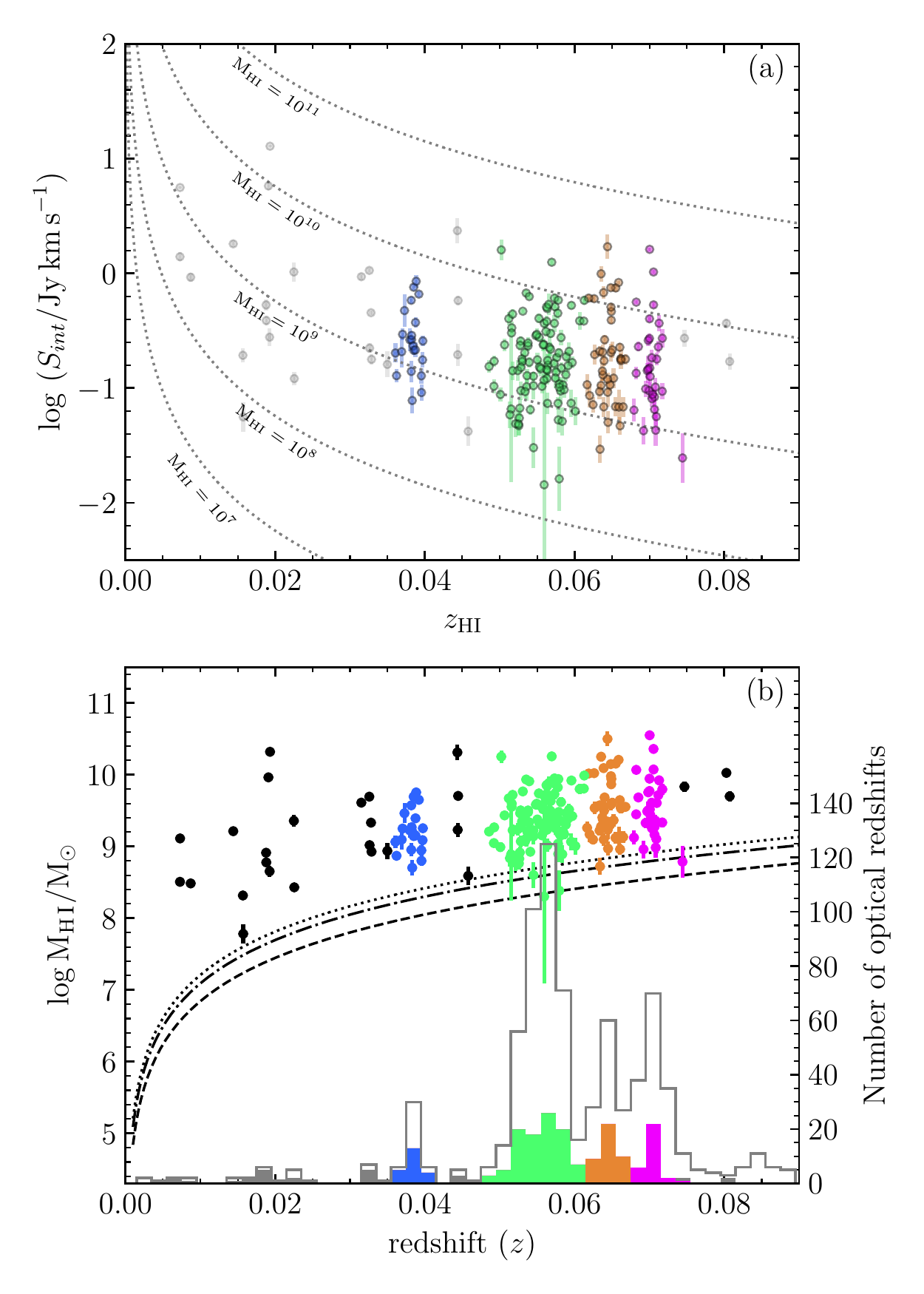}
  \caption{(a) Integrated flux vs redshift. The dotted lines show lines of constant \hi mass. (b) Measured \hi mass as a function of redshift. The coloured round data points in both panels indicate the 219 \hi detections; the colours correspond to the spectral over-densities highlighted in the histogram in panel b, and the black data points correspond to the grey histograms that include sources not associated with any over-density. The light grey open histogram represents the optical redshift catalogue from \citet{Healy2020c}. The dashed grey lines indicate the $5\sigma$ \hi mass limits at the three velocity resolutions used in the source finding: dashed for \numunit{45}{\kms}, dot-dashed for \numunit{135}{\kms}, and dotted for \numunit{225}{\kms}. }
  \label{fig:a2626_mhi_z}
 \end{figure}

As mentioned in \secref{sec:data}, the expected 3$\sigma$ \hi mass limit for a galaxy with an \hi line width of \numunit{300}{\kms} was \numunit{\mhi = 2\times 10^8}{\msun} at the distance of the cluster and the centre of the field-of-view. \figref{fig:a2626_mhi_z}b shows the distribution of the \hi masses of the detected galaxies as a function of redshift. The dashed and dotted grey lines indicate the redshift dependent mass limits at the centre of the primary beam, based on the source finding criteria.\\ 

The redshift distribution of the \hi detected galaxies is shown by the filled histogram at the bottom of \figref{fig:a2626_mhi_z}b, while the unfilled histogram represents the optical redshift catalogue covering the field from \citet{Healy2020c}. The optical catalogue is $70\%$ complete across the surveyed area for galaxies brighter than \numunit{r = 19.1}{\text{mag}}. Most of the \hi detected galaxies fall in the interval $0.0475 < z < 0.075$. {The green over-density corresponds to the redshift range of A2626 ($0.0475 < z < 0.0615$), and the orange and magenta to respectively the two background over-densities The Swarm,} an association of galaxies and galaxy groups \citep{Healy2020c} and the galaxy cluster A2637 ($0.0675 < z < 0.0745$, \citealt{Abell1989,Healy2020c}). The blue histogram identifies a distinct over-density in the foreground, likely associated with the Perseus-Pegasus filament ($0.0475 < z < 0.0615$; \citealt{Batuski1985,Healy2020c}).

%--------------------------------------------------------------------
\section{Conclusions and future work}
    \label{sec:concl}
    
    We have conducted a blind 21-cm \hi imaging survey with a single pointing of \mk in 4k mode {targeting the Abell 2626 galaxy cluster at $cz = 16576\,\kms$, over an} effective area of $\sim$ 4 deg$^{2}$ with an angular resolution of $7''.3 \times 14''.3$, a velocity resolution of 45 \kms and average noise of \numunit{80}{\mu\text{Jy/beam/channel}} over the radial velocity range of \numunit{0-25 000}{\kms}. The main summary points of this work are:
    
    \begin{enumerate}
        \item  In line with our expectations, our \mk data have enabled us to make 219 direct detections of galaxies with \hi mass \numunit{\mhi > 10^8}{\msun}. Only 12 of the galaxies were previously detected in \hi in the ALFALFA \citep{Haynes2011,Haynes2018} single-dish survey. We have presented the \hi properties of {the 219 \hi detected} galaxies as a data catalogue and as an atlas with total \hi map, \hi velocity field, signal-to-noise \hi map, position-velocity diagram {along} optical major axis, \hi global profile or spectrum, {and an} overlay with optical DECaLS r-band image for each galaxy.
        \item  We identified the \hi sources in the datacube using \sofia-2 with the smooth+clip finder algorithm with four different combinations of spatial and spectral smoothing and then identified the unique sources combining those four catalogues. The source finding was non-trivial, since the number of detections depend on determining a trade-off between \textit{reliability} and \textit{completeness}. Because of our MMT/Hectospec redshift survey and existing ancillary data from DECaLS, WINGS, SDSS, and PanSTARRS, we could finally settle at 219 \hi sources with reliable optical counterparts.
    \end{enumerate}

One of the principal objectives of observing A2626 was to image the striking jellyfish galaxy JW100 \citep{Moretti2019, Poggianti2019} and a few other jellyfish candidate galaxies (JW98, JW99, JW101, JW102, JW103, \citealt{Poggianti2015}) in \hi. Further work on the jellyfish galaxies in A2626 is presented in Deb et al. (in prep). There are three main over-densities in the line-of-sight of the observation: A2626 itself, \hwone \citep{Healy2020c} and A2637. Identification of clusters and sub-structures in this pointing area has been done and presented in \citet{Healy2020c}. {Detailed analysis of \hi morphology and kinematics of the direct \hi detections will be presented in Deb et al. (in prep) and analysis of \hi stacking of non-detections in Healy et al. (in prep)}.\\

Motivated by the unexpected high number of direct \hi detections due to MeerKAT's unprecedented sensitivity, our \hi study of A2626 will be expanded by four additional MeerKAT pointings in 32k-mode covering the ENE-SSW sector of A2626. The observations have been allocated time and will be used to investigate the \hi properties of galaxies before they have entered the cluster environment. These higher spectral resolution observations will complement the large collection of \hi detected galaxies presented in this paper, will allow detailed investigations of the kinematics of detected galaxies, and also enable the detection of low mass dwarf galaxies that escaped detection in the 4k data.

\section*{Data Availability}

Table A.1 is only available in electronic form at the CDS via anonymous ftp to \url{cdsarc.u-strasbg.fr} (130.79.128.5) or via \url{http://cdsweb.u-strasbg.fr/cgi-bin/qcat?J/A+A/}. Atlas pages (Appendix B) are available online at \url{www.astro.rug.nl/A2626/}. Reasonable requests for the \hi and continuum data presented in this work can be made to the corresponding authors.

\begin{acknowledgements}
    We thank the anonymous referee for their useful comments that have improved this work.

    We thank J Delhaize for comments on this manuscript.
    
    JH acknowledges research funding from the South African Radio Astronomy Observatory. This paper makes use of the MeerKAT data (Project ID: SCI-20190418-JH-01). The MeerKAT telescope is operated by the South African Radio Astronomy Observatory, which is a facility of the National Research Foundation, an agency of the Department of Science and Innovation. 
      
    MV acknowledges support by the Netherlands Foundation for Scientific Research (NWO) through VICI grant 016.130.338. 
    
    MR's research is supported by the SARAO HCD programme via the ``New Scientific Frontiers with Precision Radio Interferometry'' research group grant. 
    
    We acknowledge the use of the ilifu cloud computing facility -- \url{www.ilifu.ac.za}, a partnership between the University of Cape Town, the University of the Western Cape, the University of Stellenbosch, Sol Plaatje University, the Cape Peninsula University of Technology and the South African Radio Astronomy Observatory. The ilifu facility is supported by contributions from the Inter-University Institute for Data Intensive Astronomy (IDIA - a partnership between the University of Cape Town, the University of Pretoria and the University of the Western Cape), the Computational Biology division at UCT and the Data Intensive Research Initiative of South Africa (DIRISA).

    This project has received funding from the European Research Council (ERC) under the European Union’s Horizon 2020 research and innovation programme (grant agreement no. 679627; project name FORNAX)
    
    The data published here have been reduced using the CARACal pipeline, partially supported by ERC Starting grant number 679627 ``FORNAX'', MAECI Grant Number ZA18GR02, DST-NRF Grant Number 113121 as part of the ISARP Joint Research Scheme, and BMBF project 05A17PC2 for D-MeerKAT. Information about CARACal can be obtained online under the URL: \url{https://caracal.readthedocs.io}.
    
    This project has received funding from the European Research Council (ERC) under the Horizon 2020 research and innovation programme (grant agreement N. 833824). We acknowledge financial contribution from the contract ASI-INAF n.2017-14-H.0, from the grant PRIN MIUR 2017 n.20173ML3WW\_001 (PI Cimatti) and from the INAF main-stream funding programme (PI Vulcani).
    
    We acknowledge support from the Italian Ministry of Foreign Affairs and International Cooperation (MAECI Grant Number ZA18GR02) and the South African Department of Science and Innovation's National Research Foundation (DSI-NRF Grant Number 113121) as part of the ISARP RADIOSKY2020 Joint Research Scheme.
    
    This research made use of Astropy,\footnote{http://www.astropy.org} a community-developed core Python package for Astronomy \citep{Astropy2013, Astropy2018}.
      
\end{acknowledgements}

\bibliographystyle{aa}
\bibliography{references}
\appendix
\onecolumn
\section{Catalogue of \hi detections.}

\centering
\setlength{\tabcolsep}{6pt} 
\renewcommand{\arraystretch}{1.1}

\centering
\setlength{\tabcolsep}{6pt} % Default value: 6pt
\renewcommand{\arraystretch}{1.1}

\begin{table}[!h]
\caption[]{Results from the \hi synthesis observations. Only a portion of this table is shown here to demonstrate its form and content. A machine-readable version of the full table is available.}
\label{tab:catalogue}
\centering
\begin{tabular}{rcrrrrrrrrrr} \hline

\multicolumn{1}{c}{HI ID} &
\multicolumn{1}{c}{Name} &
\multicolumn{1}{c}{cz} &
\multicolumn{1}{c}{$\pm$} &
\multicolumn{1}{c}{W$_{20}$} &
\multicolumn{1}{c}{$\pm$} &
\multicolumn{1}{c}{W$_{50}$} &
\multicolumn{1}{c}{$\pm$} &
\multicolumn{1}{c}{$\int$Sdv} &
\multicolumn{1}{c}{$\pm$} &
\multicolumn{1}{c}{$F_\nu$} &
\multicolumn{1}{c}{$\pm$} \\
& & 
\multicolumn{2}{c}{km$\;$s$^{-1}$} &
\multicolumn{2}{c}{km$\;$s$^{-1}$} &
\multicolumn{2}{c}{km$\;$s$^{-1}$} &
\multicolumn{2}{c}{ Jy$\;$\kms} &
\multicolumn{2}{c}{mJy } \\

\multicolumn{1}{c}{(1)}&
\multicolumn{1}{c}{(2)}&
\multicolumn{1}{c}{(3)}&
\multicolumn{1}{c}{(4)}&
\multicolumn{1}{c}{(5)}&
\multicolumn{1}{c}{(6)}&
\multicolumn{1}{c}{(7)}&
\multicolumn{1}{c}{(8)}&
\multicolumn{1}{c}{(9)}&
\multicolumn{1}{c}{(10)}&
\multicolumn{1}{c}{(11)}&
\multicolumn{1}{c}{(12)} \\ \hline
1 & J233258.28+202618.3 & 13294 & 40 & 194 & 42 & 148 & 48 & 2.357 & 0.597 & \multicolumn{2}{c}{$<3.218$}  \\
2 & J233307.29+202331.7 & 19306 & 15 & 260 & 16 & 226 & 196 & 1.708 & 0.409 & \multicolumn{2}{c}{$<3.064$}  \\
3 & J233309.58+211415.0 & 5790 & 3 & 387 & 3 & 358 & 4 & 12.835 & 0.242 & 6.557 & 0.450  \\
4 & J233324.80+210748.4 & 5725 & 6 & 436 & 6 & 401 & 8 & 5.781 & 0.434 & 3.999 & 0.249  \\
5 & J233327.81+211747.0 & 5773 & 81 & 211 & 82 & 112 & 52 & 0.278 & 0.050 & \multicolumn{2}{c}{$<0.195$}  \\
6 & J233330.05+202112.7 & 15051 & 75 & 193 & 79 & 99 & 84 & 1.601 & 0.322 & \multicolumn{2}{c}{$<2.126$}  \\
7 & J233334.19+212112.3 & 15353 & 40 & 254 & 42 & 207 & 37 & 0.403 & 0.051 & \multicolumn{2}{c}{$<0.188$}  \\
8 & J233407.08+202522.2 & 15445 & 98 & 129 & 103 & 32 & 22 & 0.066 & 0.097 & \multicolumn{2}{c}{$<0.634$}  \\
9 & J233408.90+213128.6 & 16068 & 35 & 111 & 37 & 76 & 37 & 0.103 & 0.029 & \multicolumn{2}{c}{$<0.151$}  \\
10 & J233409.36+211641.9 & 15709 & 61 & 158 & 64 & 83 & 57 & 0.142 & 0.030 & \multicolumn{2}{c}{$<0.133$}  \\
11 & J233411.60+214044.0 & 11389 & 89 & 168 & 93 & 60 & 35 & 0.264 & 0.061 & \multicolumn{2}{c}{$<0.211$}  \\
12 & J233413.05+212327.5 & 15471 & 54 & 265 & 57 & 197 & 37 & 0.335 & 0.040 & \multicolumn{2}{c}{$<0.134$}  \\
13 & J233425.70+213122.9 & 16558 & 62 & 193 & 65 & 115 & 25 & 0.405 & 0.038 & \multicolumn{2}{c}{$<0.128$}  \\
14 & J233436.45+211302.7 & 16123 & 21 & 186 & 22 & 142 & 18 & 0.234 & 0.027 & 0.225 & 0.051  \\
15 & J233438.15+211851.7 & 16048 & 12 & 351 & 13 & 307 & 21 & 0.674 & 0.039 & 0.195 & 0.047  \\
16 & J233438.80+211721.0 & 15834 & 48 & 228 & 51 & 183 & 57 & 0.095 & 0.017 & 0.138 & 0.050  \\
17 & J233440.31+203710.8 & 17184 & 74 & 281 & 79 & 146 & 42 & 0.609 & 0.054 & 0.312 & 0.092  \\
18 & J233445.01+204514.4 & 16341 & 36 & 166 & 38 & 130 & 103 & 0.105 & 0.029 & \multicolumn{2}{c}{$<0.114$}  \\
19 & J233453.14+213344.9 & 16498 & 73 & 284 & 77 & 200 & 146 & 0.176 & 0.026 & 0.414 & 0.075  \\
20 & J233454.55+210518.6 & 17376 & 36 & 234 & 38 & 182 & 116 & 0.108 & 0.015 & \multicolumn{2}{c}{$<0.083$}  \\
21 & J233455.82+212245.8 & 15827 & 37 & 258 & 39 & 188 & 35 & 0.257 & 0.023 & \multicolumn{2}{c}{$<0.090$}  \\
22 & J233500.36+212344.3 & 19550 & 46 & 240 & 49 & 191 & 54 & 0.122 & 0.024 & 0.240 & 0.077  \\
23 & J233500.37+205908.6 & 17639 & 92 & 239 & 98 & 155 & 27 & 0.131 & 0.018 & \multicolumn{2}{c}{$<0.083$}  \\
24 & J233510.54+212147.8 & 15882 & 36 & 224 & 38 & 173 & 36 & 0.180 & 0.020 & \multicolumn{2}{c}{$<0.078$}  \\
25 & J233512.34+214628.1 & 19445 & 56 & 122 & 60 & 64 & 26 & 0.204 & 0.036 & \multicolumn{2}{c}{$<0.152$}  \\
26 & J233512.79+214138.3 & 11616 & 37 & 320 & 38 & 265 & 37 & 0.373 & 0.040 & \multicolumn{2}{c}{$<0.115$}  \\
27 & J233516.94+213314.0 & 11445 & 24 & 453 & 25 & 319 & 43 & 0.583 & 0.041 & 0.567 & 0.066  \\
28 & J233523.12+211121.2 & 22310 & 77 & 104 & 83 & 51 & 28 & 0.025 & 0.012 & \multicolumn{2}{c}{$<0.069$}  \\
29 & J233525.93+204419.5 & 18489 & 38 & 122 & 41 & 88 & 32 & 0.107 & 0.023 & \multicolumn{2}{c}{$<0.096$}  \\
30 & J233526.52+214007.5 & 11098 & 27 & 206 & 28 & 175 & 22 & 0.294 & 0.034 & \multicolumn{2}{c}{$<0.106$}  \\
31 & J233526.81+211638.3 & 19455 & 11 & 356 & 12 & 328 & 13 & 0.391 & 0.029 & 0.188 & 0.052  \\
32 & J233527.64+204059.2 & 19842 & 34 & 201 & 37 & 100 & 34 & 0.179 & 0.035 & \multicolumn{2}{c}{$<0.115$}  \\
33 & J233528.63+203803.2 & 18790 & 83 & 248 & 88 & 124 & 99 & 0.196 & 0.040 & \multicolumn{2}{c}{$<0.131$}  \\
34 & J233531.13+211524.9 & 19408 & 58 & 183 & 62 & 111 & 28 & 0.107 & 0.014 & 0.113 & 0.028  \\
35 & J233531.67+215121.8 & 18190 & 54 & 223 & 57 & 172 & 103 & 0.386 & 0.067 & \multicolumn{2}{c}{$<0.225$}  \\
36 & J233532.73+211011.3 & 19750 & 10 & 174 & 10 & 124 & 10 & 0.833 & 0.026 & 0.310 & 0.059  \\
37 & J233533.49+210252.1 & 17029 & 23 & 349 & 24 & 319 & 20 & 0.300 & 0.025 & 0.429 & 0.044  \\
38 & J233534.81+210442.8 & 16214 & 3 & 119 & 3 & 89 & 10 & 0.065 & 0.007 & \multicolumn{2}{c}{$<0.065$}  \\
39 & J233535.77+204159.6 & 18375 & 54 & 338 & 57 & 271 & 35 & 0.582 & 0.051 & 0.894 & 0.090  \\
40 & J233535.93+203807.1 & 18771 & 23 & 137 & 24 & 81 & 16 & 0.602 & 0.043 & \multicolumn{2}{c}{$<0.132$}  \\
41 & J233536.49+213059.9 & 11546 & 18 & 154 & 19 & 103 & 23 & 0.213 & 0.020 & \multicolumn{2}{c}{$<0.081$}  \\
42 & J233536.98+210440.3 & 16338 & 5 & 166 & 5 & 135 & 14 & 0.127 & 0.012 & \multicolumn{2}{c}{$<0.065$}  \\
43 & J233537.03+204639.6 & 14563 & 51 & 251 & 53 & 165 & 104 & 0.154 & 0.020 & \multicolumn{2}{c}{$<0.081$}  \\
44 & J233537.44+211025.1 & 19626 & 40 & 183 & 42 & 142 & 37 & 0.069 & 0.013 & \multicolumn{2}{c}{--}  \\
45 & J233539.99+210844.6 & 18357 & 12 & 253 & 12 & 211 & 12 & 0.384 & 0.020 & \multicolumn{2}{c}{$<0.064$}  \\
\hline
\end{tabular}
\end{table}

\newpage

\section{Examples of \hi atlas pages.}
\label{app:atlas}

\begin{figure*}[!h]
   \centering
    \includegraphics[width=1.0\textwidth]{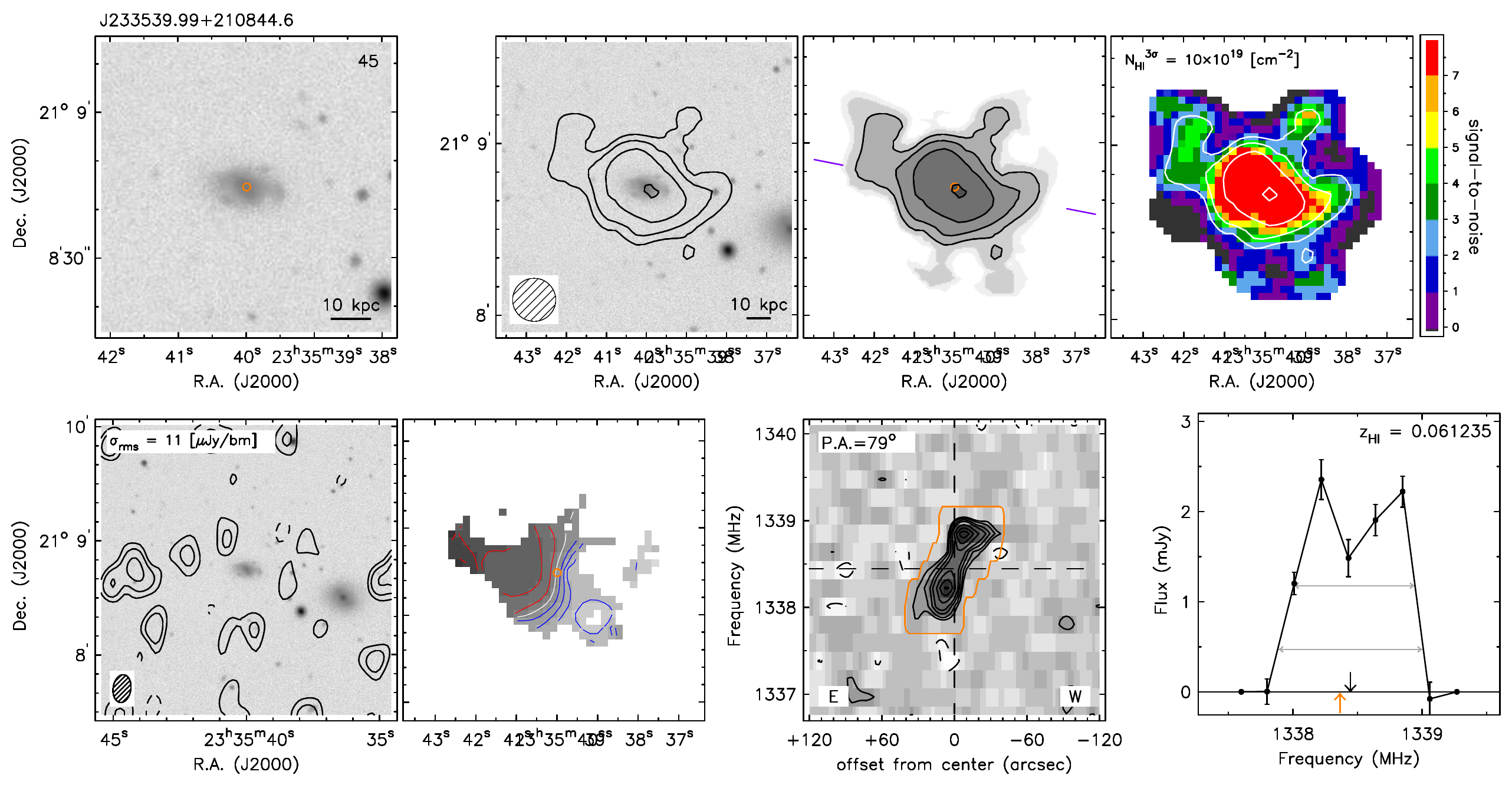}
    \vspace{-12pt}
    \caption{Atlas page for \hi ID = 45. See section \ref{sec:atlas} for details. Clockwise from the top left panel: DECaLS r-band image, total \hi map on optical image, total \hi map on greyscale, signal-to-noise map, radio continuum map on optical, \hi velocity field, position-velocity (PV) diagram, \hi global profile. The full collection of 219 atlas pages is available online.} 
\label{fig:45_atlas}
\end{figure*}

\begin{figure*}[!h]
   \centering
    \includegraphics[width=1.0\textwidth]{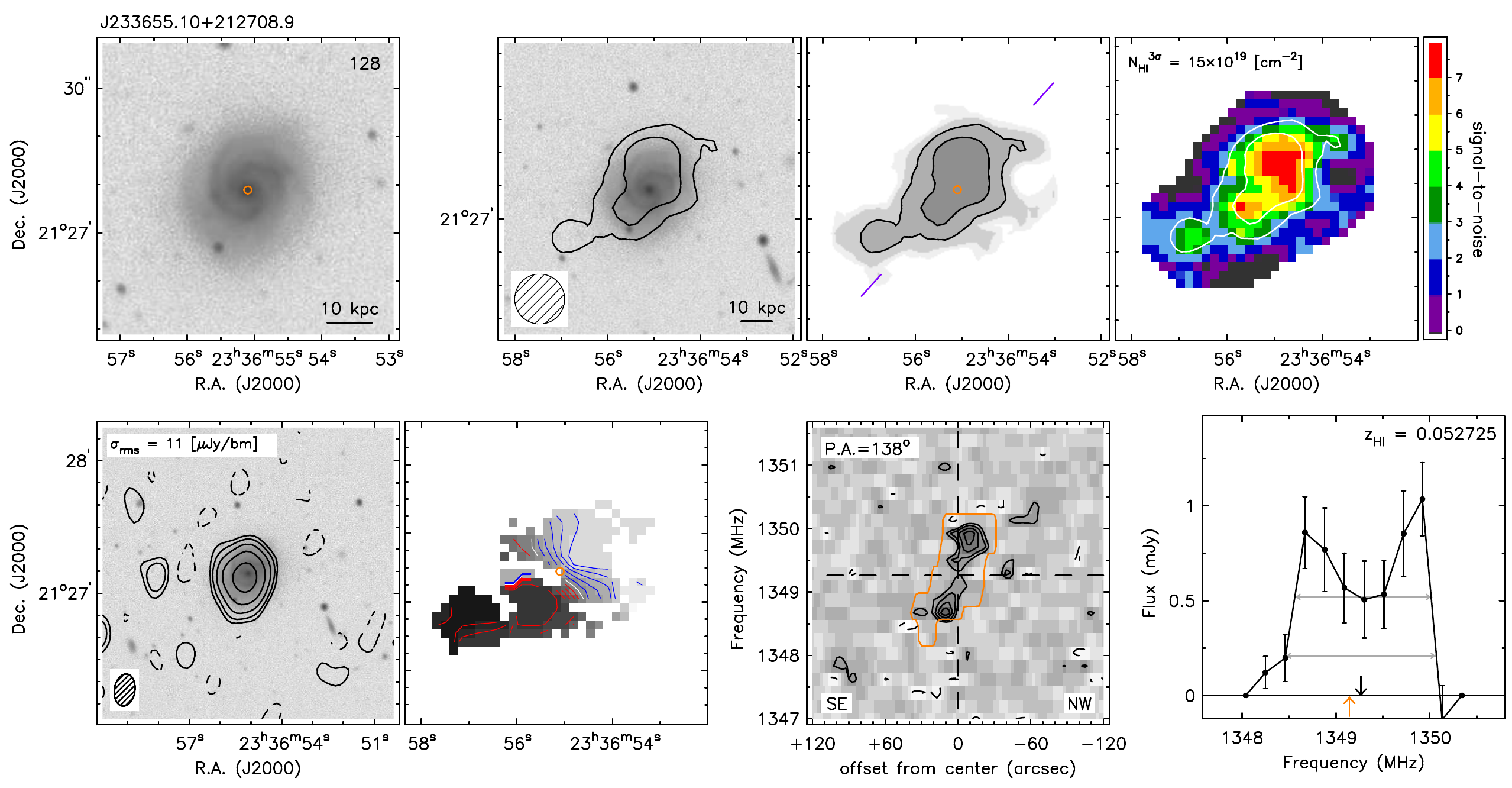}
    \caption{Atlas page for \hi ID = 128. See section \ref{sec:atlas} for details. The full collection of 219 atlas pages is available online.}
\label{fig:128_atlas}
\end{figure*}

\twocolumn

\section{SoFiA parameter file used for the\\ \hi source finding}
\label{sec:sofia_param}

{\tiny
\begin{verbatim}
### SoFiA 2.1.1 (default\_parameters.par)          ###
### Source Finding Application                     ###
### Copyright (C) 2020 Tobias Westmeier            ###

# Global settings
pipeline.verbose           =  false
pipeline.pedantic          =  true
pipeline.threads           =  0

# Input
input.data                 =  A2626.norip.cl.bs15.fits
input.region               =  
input.gain                 =  
input.noise                =  
input.weights              = 
input.mask                 = 
input.invert               =  false

# Flagging
flag.region                =  
flag.auto                  =  false
flag.threshold             =  5.0
flag.radiusSpatial         =  0
flag.log                   =  false

# Noise scaling
scaleNoise.enable          =  true
scaleNoise.mode            =  local
scaleNoise.statistic       =  mad
scaleNoise.fluxRange       =  negative
scaleNoise.windowXY        =  225
scaleNoise.windowZ         =  1
scaleNoise.gridXY          =  75
scaleNoise.gridZ           =  1
scaleNoise.interpolate     =  true
scaleNoise.scfind          =  false

# S+C finder
scfind.enable              =  true
scfind.kernelsXY           =  0
scfind.kernelsZ            =  0, 3
scfind.threshold           =  3.5
scfind.replacement         =  2.0
scfind.statistic           =  mad
scfind.fluxRange           =  negative

# Threshold finder
threshold.enable           =  false
threshold.threshold        =  5.0
threshold.mode             =  relative
threshold.statistic        =  mad
threshold.fluxRange        =  negative

# Linker
linker.radiusXY            =  3
linker.radiusZ             =  1
linker.minSizeXY           =  3
linker.minSizeZ            =  1
linker.maxSizeXY           =  0
linker.maxSizeZ            =  0
linker.keepNegative        =  false

# Reliability
reliability.enable         =  true
reliability.threshold      =  0.90
reliability.scaleKernel    =  0.15
reliability.fmin           =  5.0
reliability.plot           =  true

# Mask dilation
dilation.enable            =  false
dilation.iterations        =  1
dilation.threshold         =  -1

# Parameterisation
parameter.enable           =  true
parameter.wcs              =  true
parameter.physical         =  false
parameter.prefix           =  SoFiA
parameter.offset           =  false

# Output
output.directory           =  /data/users/deb/Cube_bs15 
output.filename            =  A2626_bs15
output.writeCatASCII       =  true
output.writeCatXML         =  false
output.writeCatSQL         =  false
output.writeNoise          =  false
output.writeFiltered       =  false
output.writeMask           =  true
output.writeMask2d         =  false
output.writeRawMask        =  false
output.writeMoments        =  true
output.writeCubelets       =  true
output.marginCubelets      =  5
output.overwrite           =  true
\end{verbatim}

}

\end{document}